\documentclass[sigconf]{acmart}

\usepackage[utf8]{inputenc} 
\usepackage[T1]{fontenc}    
\usepackage{hyperref}       
\usepackage{url}            
\usepackage{booktabs}       
\usepackage{amsfonts}       
\usepackage{nicefrac}       
\usepackage{microtype}      
\usepackage{xcolor}         
\usepackage{microtype}
\usepackage{enumitem}
\usepackage{bbm}
\usepackage{graphicx}
\usepackage{subfigure}
\usepackage{booktabs} 
\usepackage{ragged2e}
\usepackage{booktabs,makecell, multirow,tabularx}
\usepackage{algorithm}
\usepackage{algorithmic}

\usepackage{amsmath}
\usepackage{mathtools}
\usepackage{amsthm}
\usepackage{multirow}
\usepackage{diagbox}

\theoremstyle{plain}
\newtheorem{theorem}{Theorem}[section]

\theoremstyle{definition}
\newtheorem{definition}[theorem]{Definition}

\theoremstyle{remark}

\AtBeginDocument{%
  }

\copyrightyear{2025} 
\acmYear{2025} 
\setcopyright{cc}
\setcctype{by}
\acmConference[WWW '25]{Proceedings of the ACM Web Conference 2025}{April
28-May 2, 2025}{Sydney, NSW, Australia}
\acmBooktitle{Proceedings of the ACM Web Conference 2025 (WWW '25), April
28-May 2, 2025, Sydney, NSW, Australia}
\acmDOI{10.1145/3696410.3714894}
\acmISBN{979-8-4007-1274-6/25/04}
\begin{document}

\title{MER-Inspector: Assessing Model Extraction Risks from An Attack-Agnostic Perspective}

\author{Xinwei Zhang}
\affiliation{
  \institution{Hong Kong Polytechnic University}
  \city{Hong Kong}
    \country{China}
}
\email{xin-wei.zhang@connect.polyu.hk}

\author{Haibo Hu}\authornote{Corresponding author.}
\affiliation{
  \institution{Hong Kong Polytechnic University}
  \city{Hong Kong}
    \country{China}
}
\email{haibo.hu@polyu.edu.hk}

\author{Qingqing Ye}
\affiliation{%
  \institution{Hong Kong Polytechnic University}
  \city{Hong Kong}
    \country{China}
}
\email{qqing.ye@polyu.edu.hk}

\author{Li Bai}
\affiliation{%
  \institution{Hong Kong Polytechnic University}
  \city{Hong Kong}
    \country{China}
}
\email{baili.bai@connect.polyu.hk}

\author{Huadi Zheng}
\affiliation{
  \institution{Huawei Technologies Co., Ltd.}
  \city{Shenzhen}
    \country{China}
}
\email{zhenghuadi@huawei.com}

\renewcommand{\shortauthors}{Xinwei Zhang, Haibo Hu, Qingqing Ye, Li Bai, \& Huadi Zheng}

\begin{abstract}
Information leakage issues in machine learning-based Web applications have attracted increasing attention. While the risk of data privacy leakage has been rigorously analyzed, the theory of model function leakage, known as Model Extraction Attacks (MEAs), has not been well studied. In this paper, we are the first to understand MEAs theoretically from an attack-agnostic perspective and to propose analytical metrics for evaluating model extraction risks. By using the Neural Tangent Kernel (NTK) theory, we formulate the linearized MEA as a regularized kernel classification problem and then derive the fidelity gap and generalization error bounds of the attack performance. Based on these theoretical analyses, we propose a new theoretical metric called Model Recovery Complexity (MRC), which measures the distance of weight changes between the victim and surrogate models to quantify risk. Additionally, we find that victim model accuracy, which shows a strong positive correlation with model extraction risk, can serve as an empirical metric. By integrating these two metrics, we propose a framework, namely Model Extraction Risk Inspector (MER-Inspector), to compare the extraction risks of models under different model architectures by utilizing relative metric values. We conduct extensive experiments on 16 model architectures and 5 datasets. The experimental results demonstrate that the proposed metrics have a high correlation with model extraction risks, and MER-Inspector can accurately compare the extraction risks of any two models with up to 89.58\%.
\end{abstract}




\maketitle
\section{Introduction}
\label{Introduction}
Recent advancements in machine learning, particularly in deep neural networks (DNNs), have revolutionized a range of Web applications, e.g., speech recognition, image classification and sentiment analysis \cite{mcauley2012image,chen2021emoji,jia2019communitygan}.
Nonetheless, many studies have shown such machine learning-powered Web applications may cause significant information leakage issues \cite{ye2022enhanced,liu2022MLDOCTOR,zhang2023SecurityNet,liu2023provenance,huang2023training,10.1145/3704633}, which can compromise both data privacy leakage \cite{hu2024vae,10.1145/3704633,ran2024differentially} and model function leakage \cite{liu2021When}.
For model function leakage, the primary risk comes from \textit{Model Extraction Attacks (MEAs)} \cite{NEURIPS2022_4241c27d}, in which the attacker aims to copy the victim model by iteratively selecting queries using active learning techniques and then using the query-output pairs to train a surrogate model. This model extraction risk poses a threat to the intellectual property of the model owner and empowers downstream attacks, such as generating adversarial examples \cite{juuti2019PRADA}.

Given the significant leakage challenges in both model and data aspects, it is crucial for model owners to assess and mitigate the associated risks before publishing their models. While existing research primarily concentrates on data privacy risks, the analysis of model extraction risks receives less attention~\cite{liu2022MLDOCTOR,song2021Systematic}.
Although many MEAs have been proposed~\cite{NEURIPS2022_4241c27d,tramer2016Stealing, zhang2023apmsa}, their high computational costs make them unsuitable as direct measurements for model extraction risks. Furthermore, emerging advancements in MEAs inevitably lead to the underestimation of model extraction risks that are based on prior MEAs. In this paper, we aim to assess the model extraction risk from an attack-agnostic perspective because the model owner cannot know the data that the attacker has access to or the attack strategies they adopt. 

To achieve this, we first consider a worst-case threat model in current MEAs where the attacker has the union of all existing black-box MEA attackers' capabilities,  to explore the possible maximum model extraction risk. In this threat model, a powerful attacker possesses all unlabeled training samples, model architecture, initialization, and the output probabilities of the victim model without any query budget limitation~\cite{NEURIPS2022_4241c27d,pal2020ActiveThief,correia-silva2018Copycat,orekondy2019Knockoff, jagielski2020high, 10.1145/3634737.3657002}. With these capabilities, the attacker can query the victim model using all available samples and train a surrogate model, thereby achieving near-optimal attack performance~\cite{pal2020ActiveThief,orekondy2019Knockoff}.
With the aid of the Neural Tangent Kernal (NTK) theory that analyzes the non-linear neural networks by the kernel method \cite{jacot2018Neural}, we formulate this MEA as a regularized kernel classification problem that minimizes the difference in output changes between the linearized victim and surrogate models. Based on this new perspective, we derive two theoretical bounds that measure the attack performance of MEAs, namely fidelity gap (the prediction disagreement between the victim model and the surrogate model) and generalization error (the disagreement between the surrogate model's prediction and the ground truth).

Next, derived from these bounds, we propose a new theoretical attack-agnostic metric, \textit{Model Recovery Complexity (MRC)}, to assess the risk, which measures the distance between the weight change in the victim model and that in the surrogate model.  While this metric is derived from our theoretical bounds under a specific MEA, it aligns well with the model extraction risk in other practical scenarios (e.g., an attacker only has access to the output labels or has no training dataset to launch arbitrary MEAs) because it is closely related to the complexity of the model itself, as verified in Section \ref{practical_tm}. We also find that \textit{Victim Model Accuracy (VMA)}, which was known to be closely related to MEA performance~\cite{liu2022MLDOCTOR, zhang2023SecurityNet}, can serve as a second (empirical) risk metric. 
While VMA offers accurate assessments for models with large accuracy disparities, MRC captures finer details, enabling precise risk comparisons between models with similar accuracies.
Leveraging both metrics, we propose the \textit{Model Extraction Risk Inspector (MER-Inspector)}, a novel framework that effectively integrates VMA and MRC to compare risks across different models. 
Tested on 16 model architectures and 5 datasets, MER-Inspector can achieve 89.25\% accuracy in comparing the extraction risks of two models trained on the same dataset.
Our research provides model owners with crucial insights, guiding them in evaluating the efficacy of their defense strategies to ensure the deployment of more secure models.

\textbf{Contributions.} In summary, we make the following contributions:
\begin{itemize}[leftmargin=*,itemsep=0pt] 
    \item  We provide new theoretical insights into MEAs through the NTK theory, with theoretical bounds of fidelity gap and generalization error on the attack performance.
    \item We are the first to explore metrics for assessing the model extraction risk in an attack-agnostic manner and propose a theoretical metric, namely MRC, to assess this risk. 
    \item We propose the MER-Inspector framework, which combines VMA and MRC to compare the extraction risks of any two models.
    \item  We conduct extensive experiments on 16 model architectures and 5 datasets to verify the high correlation between the metrics used in MER-Inspector and model extraction risks.
\end{itemize}

The rest of this paper is organized as follows. Section \ref{related_work} introduces the background and related work. Section \ref{problem_setup} describes the problem setup. Section \ref{bound} gives theoretical bounds of MEAs, while Section \ref{Risk_Measure} details the metrics used for assessing model extraction risks. Section \ref{MP-Inspector} introduces the MER-Inspector framework. Section \ref{experiment} shows the evaluation results, followed by a discussion in Section \ref{discussion}. Finally, Section \ref{conclusion} concludes this paper.

\section{Background and Related Work}
\label{related_work}
\textbf{Model Extraction Attacks.}
MEAs, which use crafted queries to copy well-trained machine learning models illegally, pose significant threats to model intellectural property \cite{tramer2016Stealing}. Current research on model extraction goes spiral: new attacks and defenses emerge interleavingly to outwit their predecessor counterparts. For attacks, a large amount of research has been conducted to steal models in various domains efficiently, by reducing the number of queries  \cite{juuti2019PRADA,pal2020ActiveThief,correia-silva2018Copycat,orekondy2019Knockoff,yu2020CloudLeak,kariyappa2021MAZE,sanyal2022DataFree,liu2022StolenEncoder}. To counter these attacks, defense methods such as detecting extraction queries or disrupting query results have also been proposed \cite{juuti2019PRADA,zhang2023apmsa,lee2019defending,10.1007/978-3-030-29959-0_4,zheng2020protecting,zheng2020preserving,10246205}. However, no theoretical work has been done to quantify the vulnerability of models on model extraction without assuming specific attacks. Our work is the first attempt to benchmark model extraction risks from the perspective of models themselves, rather than from those attacks, thereby providing a fundamental and attack-agnostic understanding of model extraction risks.

\textbf{NTK Theory.} NTK theory is widely used to understand the training dynamics of neural networks \cite{jacot2018Neural}. 
According to NTK theory~\cite{jacot2018Neural,yang2022Informed}, the output of a wide neural network can be approximated by a linearization of its initialization using the first-order Taylor expansion:
\begin{equation}
	\label{linear}
	\mathbf{f}_{\theta}(\mathbf{x}) \approx \hat{\mathbf{f}}_\theta\left(\mathbf{x} ; \theta_{0}\right)=\mathbf{f}_{\theta_{0}}(\mathbf{x})+\bigtriangleup_\theta^{\top} \nabla_{\theta} \mathbf{f}_{\theta_{0}}(\mathbf{x}),
\end{equation}
where $\bigtriangleup_\theta = \theta - \theta_0$ is the weight change, $\theta_{0}$ is the initial weight, and $\nabla_{\theta} \mathbf{f}_{\theta_{0}}(\mathbf{x})$ represents the Jacobian of the neural network with respect to the parameters evaluated at $\theta_{0}$. The linear neural network $\hat{\mathbf{f}}_\theta(\mathbf{x};\theta_0)$ is the sum of the original output of the network and the change to the output during the training stage.
This result has been proved in infinite-width networks and further experimentally verified in ﬁnite-width networks \cite{lee2019Wide}.   Determined by empirical NTK~\cite{lee2019Wide} at $\theta_0$, the kernel function is $k(\mathbf{x},\mathbf{x'})= <\nabla_{\theta} \mathbf{f}_{\theta_{0}}(\mathbf{x}),\nabla_{\theta} \mathbf{f}_{\theta_{0}}(\mathbf{x'})> \in \mathbb{R}^{K\times K}$, and the kernel matrix $\Theta $ of training examples can be expressed as
\begin{equation}
	\label{NTK_computer}
	\Theta=\left[\begin{array}{ccc}
		k\left(\mathbf{x}^{1}, \mathbf{x}^{1}\right) & \cdots & k\left(\mathbf{x}^{1}, \mathbf{x}^{N}\right) \\
		\cdots & \cdots & \cdots \\
		k\left(\mathbf{x}^{N}, \mathbf{x}^{1}\right) & \cdots & k\left(\mathbf{x}^{N}, \mathbf{x}^{N}\right)
	\end{array}\right] \in \mathbb{R}^{N K \times N K}.
\end{equation}
In infinite-width neural networks, this NTK matrix is a positive definite matrix when no two training inputs are parallel and remains constant during training \cite{jacot2018Neural,cao2019generalization}. The NTK theory has been widely applied in various deep learning techniques, including knowledge distillation \cite{ji2020Knowledge, harutyunyan2023Supervision}, adversarial training \cite{loo2022evolution}, and neural architecture search  \cite{mok2022Demystifying}. In this work, we are the first to apply NTK theory to model extraction analysis.

\section{Problem Setup}
\label{problem_setup}
\textbf{Model.}
In this paper, we consider a classification task on the training dataset $\mathbb{D} = \{\mathbf{x}^{(i)}, y^{(i)}\}_{i=1}^{N}$, which comprises $N$ samples. Each sample is represented by a pair $(\mathbf{x}, y)$, where $\mathbf{x}$ is a feature vector in $\mathbb{R}^{d}$ with probability distribution $P_\mathbf{x}$, and $y$ is a categorical label taking on values from the set $\{1, 2, \ldots, K\}$. $d$ is the dimension of input features and $K$ is the number of labels. A deep learning model is trained on this dataset with the aim to optimize parameters $\theta \in \mathbb{R}^{p} $ that minimize a loss function $\ell$. The outputs from the final layer $\mathbf{f}_{\theta}(\mathbf{x}) =\{f_{\theta}^1(\mathbf{x}),f_{\theta}^2(\mathbf{x}),\dots,f_{\theta}^K(\mathbf{x})\}\in \mathbb{R}^{K} $ are the logits of different classes, and $y (\mathbf{x}) = \arg\max_{k \in \{1,\dots,K\}}~\mathbf{f}^k_{\theta}(\mathbf{x})$ denotes the output label of the model. As such, the output labels of the victim model and the surrogate model are $y_v(\mathbf{x})= {\arg\max}_{k \in \{1,\dots,K\}}~\mathbf{f}^k_{\theta_v}(\mathbf{x})$ and $y_s(\mathbf{x})= {\arg\max}_{k \in \{1,\dots,K\}}~\mathbf{f}^k_{\theta_s}(\mathbf{x})$, respectively.

\textbf{Model Extraction Risk.}  
We consider the model extraction risk from state-of-the-art MEAs \cite{NEURIPS2022_4241c27d,pal2020ActiveThief,correia-silva2018Copycat,orekondy2019Knockoff} where the attacker iteratively selects query samples using active learning techniques to minimize the number of queries. The objectives of MEAs includes attack accuracy and fidelity \cite{jagielski2020high}. The former refers to the attacker's intention to replicate the ground truth value. This objective is to keep the surrogate model performing well on the testing dataset. The latter refers to the attacker's attempt to replicate the victim model's output. This ensures the surrogate model can replace the victim model for downstream attacks, such as adversarial examples. 
In most MEA works, the latter one is considered more important \cite{jagielski2020high}, which we define it as follows:
\begin{definition}[Fidelity]
    Let $\mathbf{f}_{\theta_v}$ be a victim model. Given the target distribution $P_\mathbf{x}$ over $\mathbf{x}$, and a similarity function $S(\mathbf{f}_{\theta_v}, \mathbf{f}_{\theta_s}) = \mathbbm{1}(\arg \max \mathbf{f}_{\theta_s} = \arg \max \mathbf{f}_{\theta_v})$, the fidelity $\mathbbm{F}(\mathbf{f}_{\theta_s},\mathbf{f}_{\theta_v})$ between the victim model and a surrogate model $\mathbf{f}_{\theta_s}$, where $\mathbf{f}_{\theta_s}$ is constructed through a MEA, is defined as the probability $\Pr_{\mathbf{x} \sim P_\mathbf{x}}[S(\mathbf{f}_{\theta_v}(\mathbf{x}), \mathbf{f}_{\theta_s}(\mathbf{x}))]$.

\end{definition}

A specific MEA $\mathcal{A}(\mathcal{C}, \mathcal{S})$ depends on the attackers' capabilities $\mathcal{C}$ and strategies $\mathcal{S}$. Under a specific MEA, the model extraction risk can be defined as follows.

\begin{definition}[Model Extraction Risk]
	Let $\mathbf{f} _{\theta _v}$ be the victim model and $\mathcal{A}(\mathcal{C}, \mathcal{S})$ be a MEA. The extraction risk of the victim model under $\mathcal{A}(\mathcal{C}, \mathcal{S})$ is defined as the fidelity the attack can achieve, i.e., 
	\begin{equation}
		\text{Risk}(\mathbf{f} _{\theta _v}) : = \max _{\mathbf{f} _{\theta _s} \in \mathcal{F} _{\mathcal{A}}(\mathcal{C} ,\mathcal{S})}  \mathbbm{F}(\mathbf{f}_{\theta_s},\mathbf{f}_{\theta_v}) ,
	\end{equation}
	where $\mathcal{F} _{\mathcal{A}}(\mathcal{C} ,\mathcal{S})$ denotes the set of surrogate models $\mathbf{f} _{\theta _s}$ constructed through the attack  $\mathcal{A}(\mathcal{C}, \mathcal{S})$. Particularly, observations of increasing fidelity performance the attack can achieve, the higher model extraction risk of the victim model.
\end{definition}

\textbf{Threat Model.} From the perspective of the model owner, they cannot predict the specific MEA method that the attacker may adopt. Therefore, in an attacker-agnostic manner, this paper explores the maximum possible model extraction risk under a worst-case threat model, wherein the attacker possesses the union of all existing black-box MEA attackers' capabilities. These capabilities, defined as the attackers' maximum capabilities $\mathcal{C}_m$, include (1) full access to the victim model outputs, i.e., the full posterior; (2) knowledge of all unlabeled training samples, model architecture, and initialization; and (3) no limitations on the query budget \cite{NEURIPS2022_4241c27d,pal2020ActiveThief,correia-silva2018Copycat,orekondy2019Knockoff, jagielski2020high, 10.1145/3634737.3657002}. In this case, the attacker can query all training samples and achieve nearly maximum fidelity without considering any specific attack strategies \cite{pal2020ActiveThief,orekondy2019Knockoff}.
The risk posed by $\mathcal{A}(\mathcal{C}_m, \mathcal{S})$ can thus be regarded as the model extraction risk from the model owner's perspective.
Despite the attacker's formidable capabilities, they still cannot perfectly replicate the victim model through retraining due to the lack of access to ground-truth labels and the randomness in training variables like hyperparameters and execution environments~\cite{ahn2022reproducibility}. 

\textbf{Linearized Model Extraction.} 
This paper formulated the MEA under the above threat model as linearized model extraction. We assume the victim and surrogate models are over-parameterized and wide. This assumption enables us to theoretically analyze MEAs by applying NTK theory, which allows the substitution of nonlinear neural networks with linear ones.
The functions of the victim model and the surrogate model can be formulated as $\mathbf{f}_{\theta_v}(\mathbf{x}) \approx \mathbf{f}_{\theta_0}(\mathbf{x}) + \bigtriangleup_{\mathbf{f}_{\theta_v}(\mathbf{x})} = \mathbf{f}_{\theta_{0}}(\mathbf{x})+\bigtriangleup_{\theta_v}^{\top} \nabla_{\theta} \mathbf{f}_{\theta_{0}}(\mathbf{x})$ and $\mathbf{f}_{\theta_s}(\mathbf{x}) \approx \mathbf{f}_{\theta_0}(\mathbf{x}) + \bigtriangleup_{\mathbf{f}_{\theta_s}(\mathbf{x})} =  \mathbf{f}_{\theta_{0}}(\mathbf{x})+\bigtriangleup_{\theta_s}^{\top} \nabla_{\theta} \mathbf{f}_{\theta_{0}}(\mathbf{x})$, respectively. Here the terms $\mathbf{f}_{\theta_{0}}(\mathbf{x})$ and $\nabla_{\theta} \mathbf{f}_{\theta_{0}}(\mathbf{x})$ only depend on the model initialization. 
For simple analysis, we assume the same initialization parameters to focus on analyzing the MEA process. Consequently, the MEA for linearized neural networks can be formulated by a regularized kernel classification problem, i.e.,
\begin{equation}
\label{optimization}
\bigtriangleup_{\mathbf{f}_{\theta_s}(\mathbf{x})}^{*} \in \underset{\bigtriangleup_{\mathbf{f}_{\theta_s}(\mathbf{x})} \in \mathcal{H}}{\operatorname{argmin}} \frac{1}{N} \sum_{i=1}^{N} \ell\left(\bigtriangleup_{\mathbf{f}_{\theta_s}(\mathbf{x})}^{(i)}, \bigtriangleup_{\mathbf{f}_{\theta_v}(\mathbf{x})}^{(i)}\right) +\frac{\lambda}{2}\|\bigtriangleup_{\mathbf{f}_{\theta_s}(\mathbf{x})}\|_{\mathcal{H}}^{2},
\end{equation}
where $\ell$ is the loss function and $\|\bigtriangleup_{\mathbf{f}_{\theta_s}(\mathbf{x})}\|_{\mathcal{H}}$ is the Reproducing Kernel Hilbert Space (RKHS) norm of the function $\bigtriangleup_{\mathbf{f}_{\theta_s}(\mathbf{x})}$. Based on the representer theorem \cite{scholkopf2001generalized}, the solution to this problem can be expressed as
\begin{equation}
\label{F_change}
	\bigtriangleup_{\mathbf{f}_{\theta_s}(\mathbf{x})} =\sum_{i=1}^{N}\alpha_i k(\mathbf{x},\mathbf{x}^{(i)})= \Theta \boldsymbol{\alpha},
\end{equation}
and the RKHS norm of $\bigtriangleup_{\mathbf{f}_{\theta_s}(\mathbf{x})}$ can be expressed as $\|\bigtriangleup_{\mathbf{f}_{\theta_s}(\mathbf{x})}\|_{\mathcal{H}}^{2}=\sum_{i=1}^{N} \sum_{j=1}^{N} \alpha_{i} \alpha_{j} k\left(x^{(i)}, x^{(j)}\right)=\boldsymbol{\alpha}^{\top} \Theta \boldsymbol{\alpha}$,
where $\alpha_i$ is the weight of the corresponding kernel function $k(\mathbf{x},\mathbf{x}_i)$.

Based on the regularized kernel classification problem, we can analyze the attack performance that an attacker achieves by solving this problem. In the following sections, we will analyze the attack performance bounds of linearied MEAs and propose theoretical risk metrics to bridge the gap between these attack-related performance bounds and practical attack-agnostic evaluations.

\section{Theoretical Bounds of MEA}
\label{bound}

\textbf{Fidelity Gap Bound.}
The fidelity gap is the probability of the prediction disagreement between the surrogate model and the victim model.
Let us define the prediction margin $\mathcal{M}(\mathbf{x},y_v,y_s) = f_{\theta_s}^{y_v}(\mathbf{x}) - \max_{y'\ne {y_v}}f_{\theta_s}^{y'}(\mathbf{x})$, and the fidelity gap $\mathcal{G} = \underset{\mathbf{x} \sim P_{\mathbf{x}}}{\mathbb{P}}\left[y_s(\mathbf{x}) \neq y_v(\mathbf{x})\right] = \underset{\mathbf{x} \sim P_{\mathbf{x}}}{\mathbb{P}}\left[ \mathcal{M}(\mathbf{x},y_v,y_s) \le 0\right]$.
Since $\bigtriangleup_{{\mathbf{f}}_{\theta_s}(\mathbf{x})}$ exists in the RKHS $\mathcal{H}$ corresponding to the kernel $k$, the fidelity gap can be upper bounded by the following theorem and its proof is available in Appendix \ref{proof4.3}. 
\begin{theorem}[Fidelity gap bound]
\label{fidelity_gap_bound}
        Assume that \(\kappa = \sup_{\mathbf{x} \sim P_{\mathbf{x}}} k(\mathbf{x},\mathbf{x}) < \infty\). Define a constant \(\gamma > 0\), and let \(M_{0} = \left\lceil\frac{\gamma \sqrt{N}}{4 K \sqrt{\kappa}}\right\rceil\). Subsequently, with a probability of at least \(1 - \delta\), for every function \(\bigtriangleup_{\mathbf{f}_{\theta_s}(\mathbf{x})} \in \mathcal{H}\), the fidelity gap on $N$ training samples can be bounded:
	\begin{equation}
		\begin{aligned}
			\mathcal{G}_N = & ~\underset{\mathbf{x} \sim P_{\mathbf{x}}}{\mathbb{P}} [ \mathcal{M}(\mathbf{x},y_v,y_s) \leq 0 ] \\ \leq
			& \frac{1}{N} \sum_{i=1}^{N} \mathbf{1}\left\{\mathcal{M}(\mathbf{x}^{(i)},y_v^{(i)},y_s^{(i)}) \leq \gamma\right\}\\
			&+\frac{4 K(\bigtriangleup^{\top}_{{\mathbf{f}}_{\theta_v}(\mathbf{x})}\Theta^{-1}\bigtriangleup_{{\mathbf{f}}_{\theta_v}(\mathbf{x})})}{\gamma N} \sqrt{\text{Tr} (\Theta)}
			+3 \sqrt{\frac{\log \left(2 M_{0} / \delta\right)}{2 N}},
		\end{aligned}
	\end{equation}
	where $\text{Tr} (\Theta)$ represents the trace of the victim model's NTK matrix $\Theta$.
\end{theorem}

The bound consists of three main components: The first term is an empirical margin loss based on the training data, which is smaller than 1 even when the $\gamma$ is infinite.
The second term is a penalty term related to the model complexity, which means that a more complex victim model leads to a large fidelity gap. The margin parameter $\gamma$ influences the first two terms, which need to be chosen carefully to control the balance between the empirical margin loss and the victim model complexity~\cite{harutyunyan2023Supervision}. The third item is the sample complexity item, which is related to the selected confidence $\delta$, and represents the reliability of the generalization error bound under a certain confidence level.


\textbf{Generalization Error Bound.}
The generalization error of the surrogate model is the probability of disagreement between the prediction of the surrogate model and the ground truth, i.e., $\mathcal{R}^s = \underset{\mathbf{x} \sim P_{\mathbf{x}}}{\mathbb{P}}\left[y_s(\mathbf{x}) \neq y(\mathbf{x})\right]$. Similarly, the generalization error of the victim model $\mathcal{R}^v = \underset{\mathbf{x} \sim P_{\mathbf{x}}}{\mathbb{P}}\left[y_v(\mathbf{x}) \neq y(\mathbf{x})\right]$. The following theorem proves that the generalization error of the surrogate model is bounded by the generalization performance of the victim model and the fidelity gap between them, and its proof is available in Appendix \ref{proof4.4}. 

\begin{theorem}[Generalization risk bound]
\label{Generalization_risk_bound}
	Given the fidelity gap bound \( \mathcal{G}_N\) and the victim model's generalization error bound \(\mathcal{R}_N^v\). Then, the surrogate model's generalization error is bounded by,
	\begin{equation}
		\begin{aligned}
			\mathcal{R}_N^s &\leq \mathcal{G}_N + \mathcal{R}_N^v. \\
		\end{aligned}
	\end{equation}
\end{theorem}




\section{Model Extraction Risk Measure}
\label{Risk_Measure}
Until now, existing evaluation of model extraction risks against MEAs has to be empirical, i.e., by conducting a range of well-known MEAs~\cite{NEURIPS2022_4241c27d,tramer2016Stealing, zhang2023apmsa} and measure the real fidelity and accuracy. However, this approach is neither cost-effective nor future-proof. Additionally, directly calculating the attack performance bound of Theorem \ref{fidelity_gap_bound} to assess risk is impractical due to the high computational cost, impractical threat model and the need for careful parameter $\gamma$ selection. In this section, we explore attack-agnostic metrics that can precisely capture the risk associated with the victim model.
We first introduce an empirical metric, VMA, for a rough risk assessment, followed by a theoretical metric, MRC, derived from the fidelity gap bound of Theorem \ref{fidelity_gap_bound}, for a more fine-grained assessment.
\begin{figure}[t]
\begin{center}
\centerline{\includegraphics[width=0.7\columnwidth]{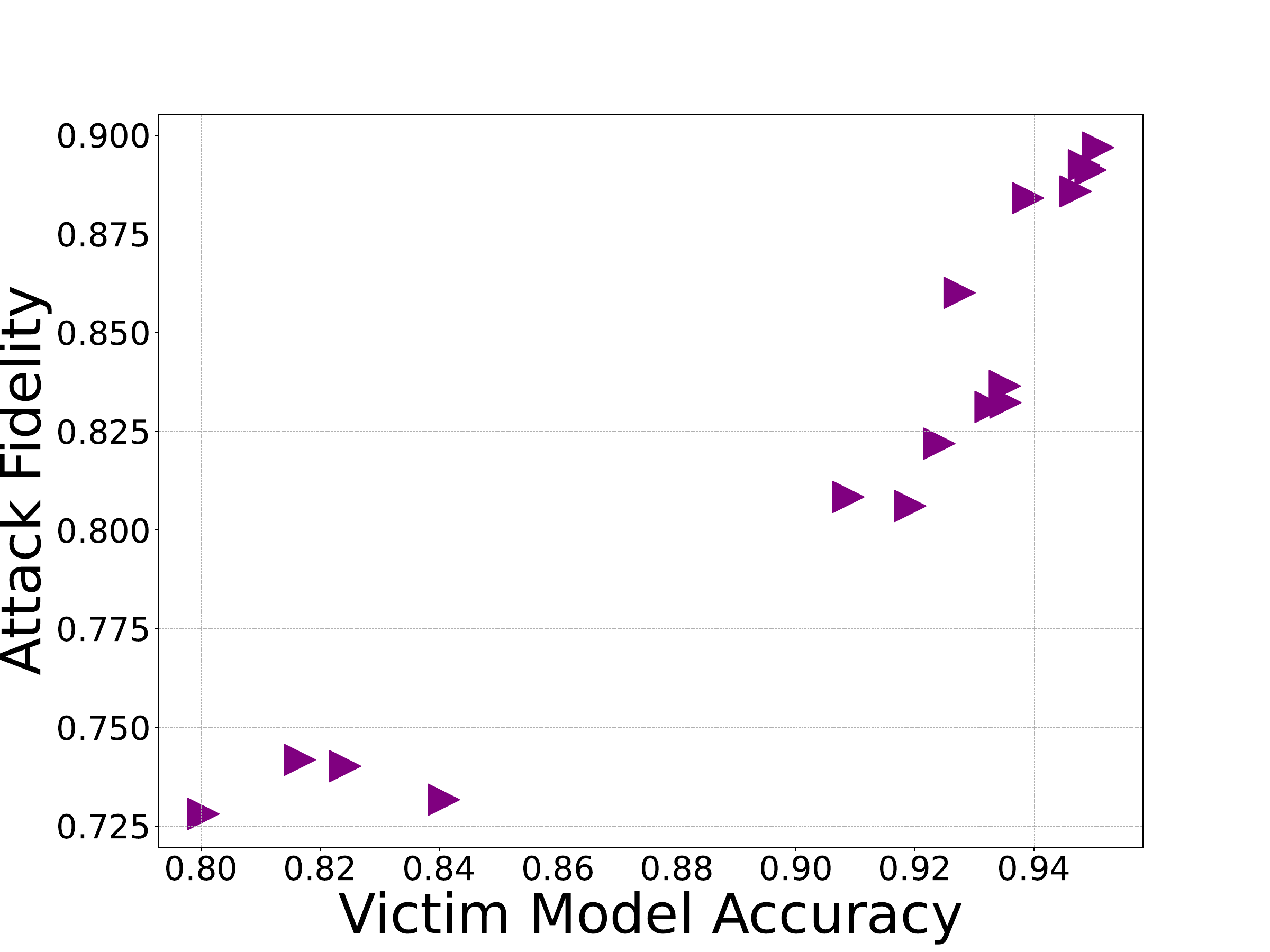}}
\caption{Relationship between attack fidelity and VMA.}
\label{VMA_AF}
\end{center}
\end{figure}

\subsection{Victim Model Accuracy}
The first term of empirical loss in Theorem \ref{fidelity_gap_bound} directly reflects the model extraction risk experimentally, yet it is dependent on a specific MEA. To establish an attack-agnostic measurement, we find that the empirical VMA effectively indicates attack fidelity. This enables us to assess potential risks without being limited by the particulars of any MEA. \citet{zhang2023SecurityNet} have experimentally demonstrated a significant correlation between VMA and both attack accuracy and fidelity. Our empirical findings further confirm the strong relationship between VMA and attack fidelity, reinforcing the utility of VMA as an attack-agnostic metric in risk assessments.
As shown in Figure \ref{VMA_AF}, we observe a strong positive relationship exists between them, with a Pearson Correlation Coefficient (PCC) of 0.9364 and a Kendall Rank Correlation Coefficient (KRC) of 0.85. However, we argue that VMA can only provide a rough risk measurement and may not effectively indicate the risk when the accuracies of these compared models are nearly identical, e.g., both close to 95\% (see Section~\ref{results} and Appendix~\ref {Effect_Metircs_Datasets}).
Furthermore, VMA is an empirical metric without a robust theoretical foundation to support it. Therefore, we need a more fine-grained and theoretical metric to measure the risk.

\subsection{Model Recovery Complexity}
Although Theorem \ref{fidelity_gap_bound} provides a theoretical attack performance bound of a specific MEA rather than in an attack-agnostic manner, it shows that the attack performance is highly related to the victim model's complexity. Specifically, in Theorem \ref{fidelity_gap_bound}, the expression \(\bigtriangleup^{\top}_{{\mathbf{f}}_{\theta_v} (\mathbf{x})} \Theta^{-1} \bigtriangleup_{\mathbf{f}_{\theta_v}(\mathbf{x})}\) in the second term can reflects the model complexity and has a clear and intuitive explanation that we elaborate on later. Additionally, this expression bears similarity to NTK-based metrics proposed in previous work, such as label-gradient alignment $Y^T \Theta Y$~\cite{mok2022Demystifying} and supervision complexity $Y^T \Theta^{-1} Y$~\cite{harutyunyan2023Supervision}, whose inherent formula enables it to capture a large number of nonlinear features. Consequently, we formally define it as MRC as follows.

\begin{definition}[Model Recovery Complexity (MRC)]
	Given the victim model $\mathcal{F}$ and training dataset $\mathbb{D} =\{\mathbf{x}^{(i)},y^{(i)}\}_{i=1}^N$, the MRC of model $\mathcal{F}$ is
	\begin{equation}
		r_{rc}(\mathbb{D},\mathcal{F}) = \bigtriangleup^{\top}_{{\mathbf{f}}_{\theta_v}(\mathbf{x})}\Theta^{-1}\bigtriangleup_{{\mathbf{f}}_{\theta_v}(\mathbf{x})},
		\label{r1}
	\end{equation}
	where $\bigtriangleup_{{f}_{\theta_v}(\mathbf{x})}$ denotes the output change of the victim model before and after model training on all training samples $\mathbf{x}$, and $\Theta$ is the kernel matrix defined by NTK.
\end{definition}
In essence, $\Theta$ represents a mapping of data into a feature space defined by the neural network, and metric MRC quantifies the "\textit{size}" or "\textit{impact}" of output changes in this feature space.
To better understand MRC, let us assume the weight change of the victim model is $\bigtriangleup_{\theta_v} = \theta_v - \theta_0$, and the weight change of the surrogate model is $\bigtriangleup_{{\theta_s}} = \theta_s - \theta_0$.
Let $\mathbf{X}= [\mathbf{x}_1,\mathbf{x}_2,...,\mathbf{x}_M] \in \mathbb{R}^{d\times M}$ and $\mathbf{Y}=[\mathbf{f}_{\theta_v}(\mathbf{x}_1),\mathbf{f}_{\theta_v}(\mathbf{x}_2),...,\mathbf{f}_{\theta_v}(\mathbf{x}_M)] \in \mathbb{R}^{K\times M}$ denote the thief dataset. According to Theorem 1 of \cite{phuong2019Understanding}, the converged weight change of the surrogate model is
\begin{equation}
	\begin{aligned}
		\Delta_{{\theta_s},M}
		&=\nabla_{\theta_v,M} \mathbf{f}_{\theta_{0}}(\mathbf{x})~\Theta^{-1}~\Delta_{f_{\theta_v,M}(\mathbf{x})} \\
		&=\nabla_{\theta_v,M} \mathbf{f}_{\theta_{0}}(\mathbf{x})~\Theta^{-1}~\nabla^\top_{\theta_v,M} \mathbf{f}_{\theta_{0}}(\mathbf{x}) ~\Delta_{\theta_v}
		 =\mathbf{P}_{\theta_v,M}~ \Delta_{\theta_v},
	\end{aligned}
\end{equation}
where $\mathbf{P}_{\theta_v,M}$ is a projection matrix onto span $\{\nabla_{\theta_v} \mathbf{f}_{\theta_{0}}(\mathbf{x})\}_M$. According to this equation, the weight change learned by the attacker is simply the projection of the victim model's weight change onto the span $\{\nabla_{\theta_v} \mathbf{f}_{\theta_{0}}(\mathbf{x})\}_M$. As $M$ increases and span $\{\nabla_{\theta_v} \mathbf{f}_{\theta_{0}}(\mathbf{x})\}_M$ expands, $\bigtriangleup_{{\theta_s}}$ gradually converges to $\bigtriangleup_{\theta_v}$.
Therefore, MRC can be rewritten as
\begin{equation}
	\begin{aligned}
		r_{rc}(\mathbb{D},\mathcal{F})
		&=\bigtriangleup^{\top}_{{f}_{\theta_v}(\mathbf{x})}\Theta^{-1}\bigtriangleup_{{f}_{\theta_v}(\mathbf{x})}\\
	&	=\bigtriangleup^{\top}_{\theta_v} \nabla_{\theta_v} \mathbf{f}_{\theta_{0}}(\mathbf{x}) \Theta^{-1} \nabla^{\top}_{\theta_v} \mathbf{f}_{\theta_{0}}(\mathbf{x})\bigtriangleup_{\theta_v} 
		= \bigtriangleup^{\top}_{\theta_v} \mathbf{P}_{\theta_v} \bigtriangleup_{\theta_v}.
	\end{aligned}
\end{equation}
Essentially, the metric calculates the "\textit{distance}" between the projected vector $\mathbf{P}_{\theta_v} \bigtriangleup_{\theta_v}$ and the original vector $\bigtriangleup_{\theta_v}$.
If the metric value is small, it means that the projected vector is close to the original vector.
As such, the difference in weight changes between the victim model and the surrogate model is small, thereby increasing the model extraction risk.

\subsection{MRC Approximation}


Although MRC can measure the risk, accurately computing its score presents two challenges.
\begin{itemize}
    \item Time-consuming. The time complexity of computing the NTK matrix is $O((NK)^2p)$ \cite{pmlr-v162-novak22a}. For example, on CIFAR10, it takes over 1,000 GPU hours to compute the NTK matrix.
    \item Inaccuracy. The NTK matrix is accurate in infinite-width neural networks. But when we apply it to commonly used networks such as ResNet \cite{he2016deep} and DenseNet \cite{huang2017densely}, the NTK-based metric may not align perfectly with our theoretical expectations.
\end{itemize}

To address the first challenge, we perform sampling to reduce the number of samples to calculate the NTK matrix.
To ensure that the selected samples yield more accurate approximations, it is essential to answer a key question: \textbf{which samples are crucial to dictate MEA performance?}


\begin{figure}[t]
    \centering
    \subfigure[JBDA]{
        \includegraphics[width=0.22\textwidth]{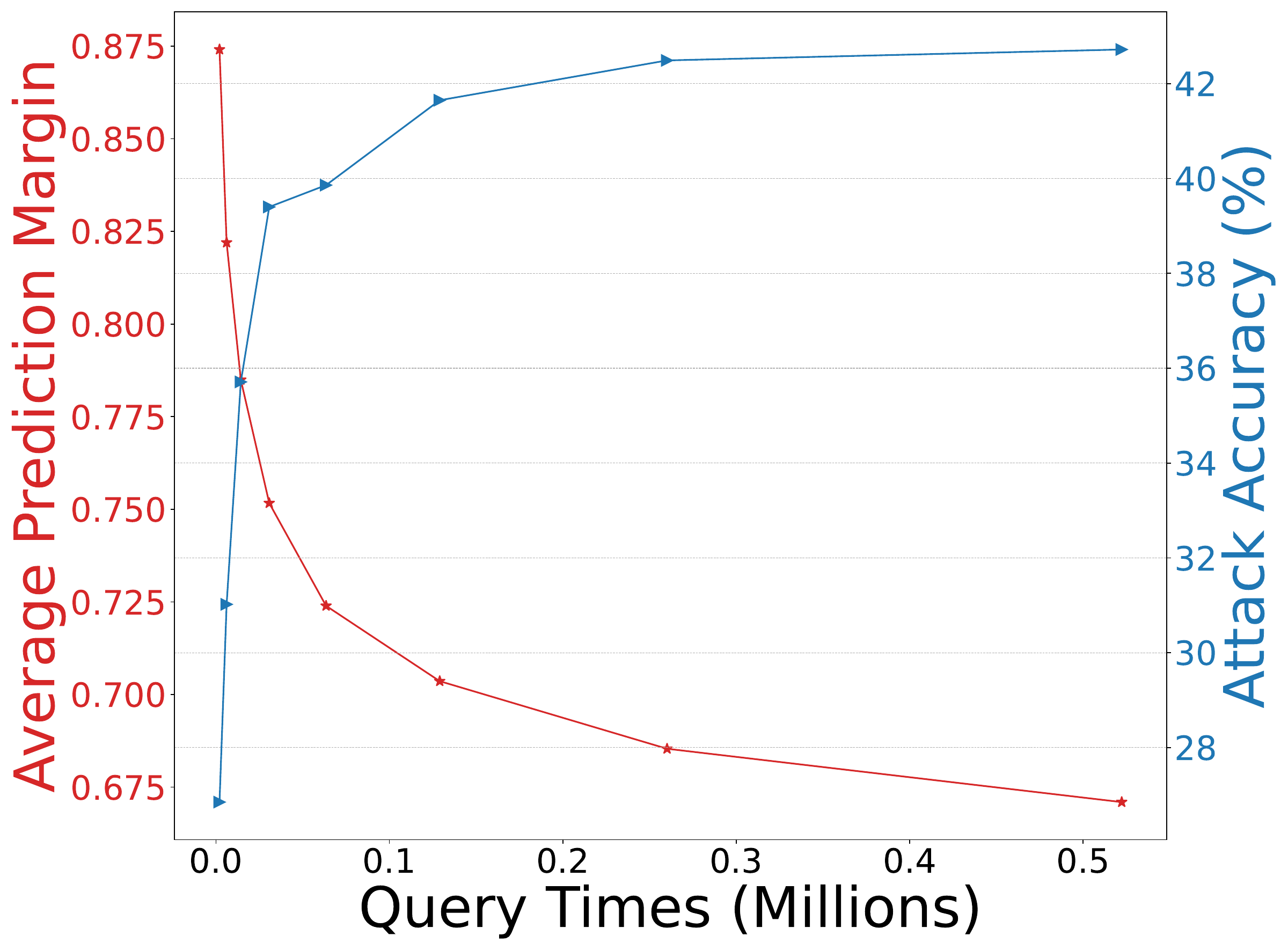}
        \label{fig:Margin_JBDA}
    }
    \subfigure[MAZE]{
        \includegraphics[width=0.22\textwidth]{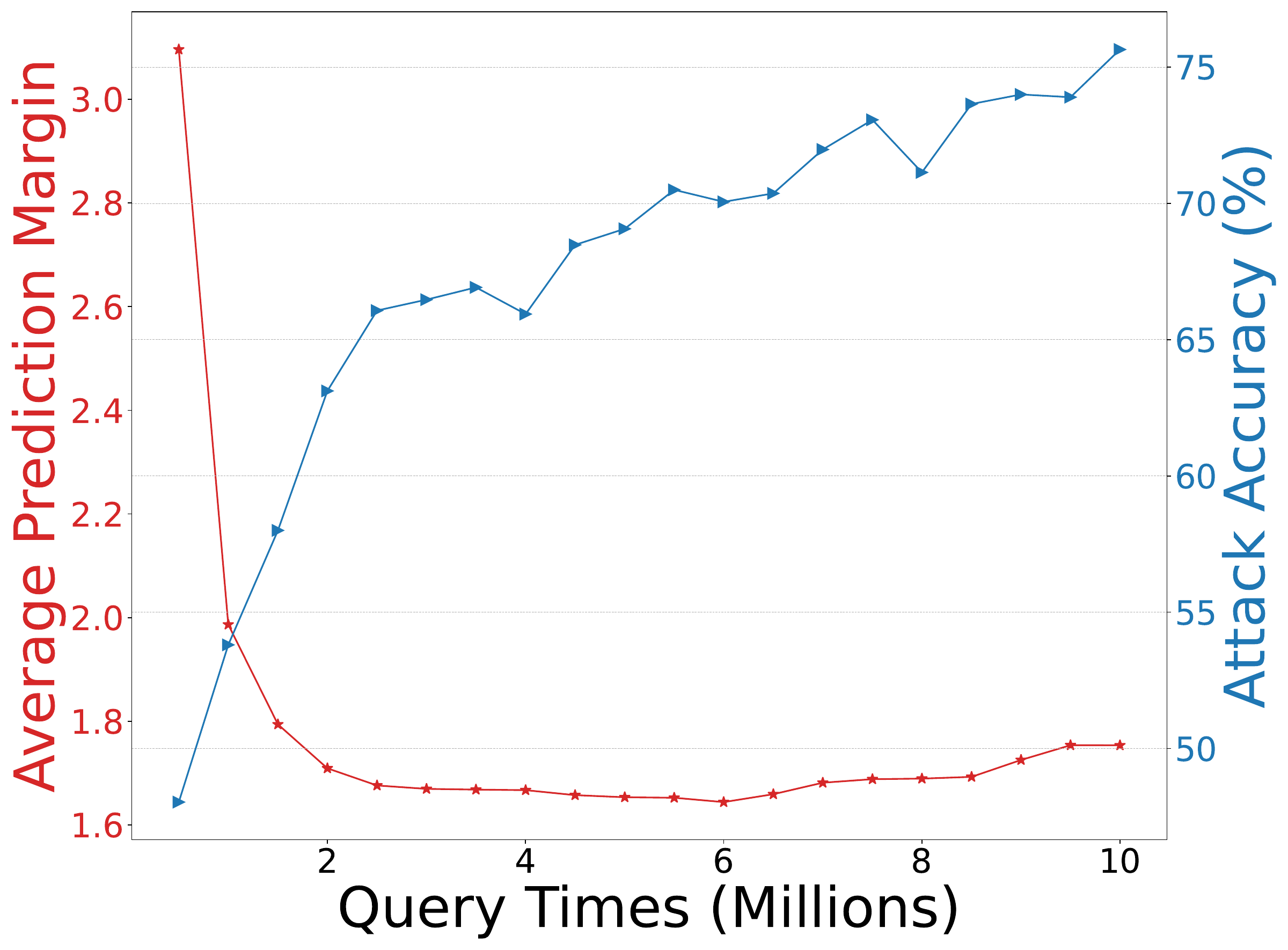}
        \label{fig:Margin_MAME}
    }
    \caption{The change of average prediction margin and attack accuracy with increased query times under (a) JBDA, and (B) MAZE on CIFAR 10.}
    \label{fig:Margin}
\end{figure}

We conducted two well-known attacks, JBDA \cite{papernot2017Practical} and MAZE \cite{kariyappa2021MAZE}, to analyze the variations in the average prediction margin and attack accuracy as the number of queries increases. In each iteration, these two methods use the surrogate model held at this time to synthesize query samples for improving query efficiency. The experiment details are given in Appendix \ref{JDBA_MAZE_experiment}. Figure \ref{fig:Margin} shows the trend that the average prediction margin of query samples decreases with increased attack performance. This means that the attacker first uses simple samples (with larger prediction margins) to obtain a rough classification boundary, and then gradually finds hard samples (with small prediction margins) to refine the classification boundary. \citet{sorscher2022Neural} also highlighted that model training benefits the most from both simple and hard samples. Inspired by these findings, we choose those samples with the largest and smallest margins as both are crucial to attack performance. Specifically, among a total of $L$ samples for MRC approximation, we select $L_d$ samples with the largest margin and $L-L_d$ samples with the smallest margin, where $\eta = L_d / L$ is the \textit{difficulty ratio}.

To address the second challenge, we find that most NTK matrices calculated in finite-width neural networks do not conform to the theoretical expectations. First, it cannot remain constant during training. \citet{mok2022Demystifying} demonstrated that an NTK matrix changes significantly at the beginning of the training and remains constant in subsequent training. Therefore, to get the constant NTK matrix, unlike other works that compute NTK-based metrics at the initialization stage \cite{mok2022Demystifying,xu2021knas}, we compute the NTK matrix after the training. Second, the computed NTK matrix is not always positive definite, so its inversion $\Theta^{-1}$ is not stable. We solve this by adjusting the minimum eigenvalue of the NTK matrix to a threshold $q$. In addition, in order to maintain a certain stability in the MRC score, we use the probabilities of the model output instead of logits in the calculation. The overall process of MRC approximation algorithm is shown in Appendix \ref{A1}.


\section{MER-Inspector: A Risk Comparison Framework}
\label{MP-Inspector}

From the above analysis, the VMA and MRC of the victim model can serve as good metrics for evaluating the model extraction risk. However, either VMA or MRC has its own limitations. The VMA is less effective in assessing the risk disparity between models with insignificant variation in accuracy. MRC leverages the NTK theory, whose quantitative analysis may be inconsistent with the actual behavior of finite-width networks. Nonetheless, we find MRC achieves good relative accuracy, i.e., when comparing models of similar structures (e.g., in the ResNet family) even if their accuracies differ little. 

\begin{figure}[t]
	\centering
	\includegraphics[width=0.9\columnwidth]{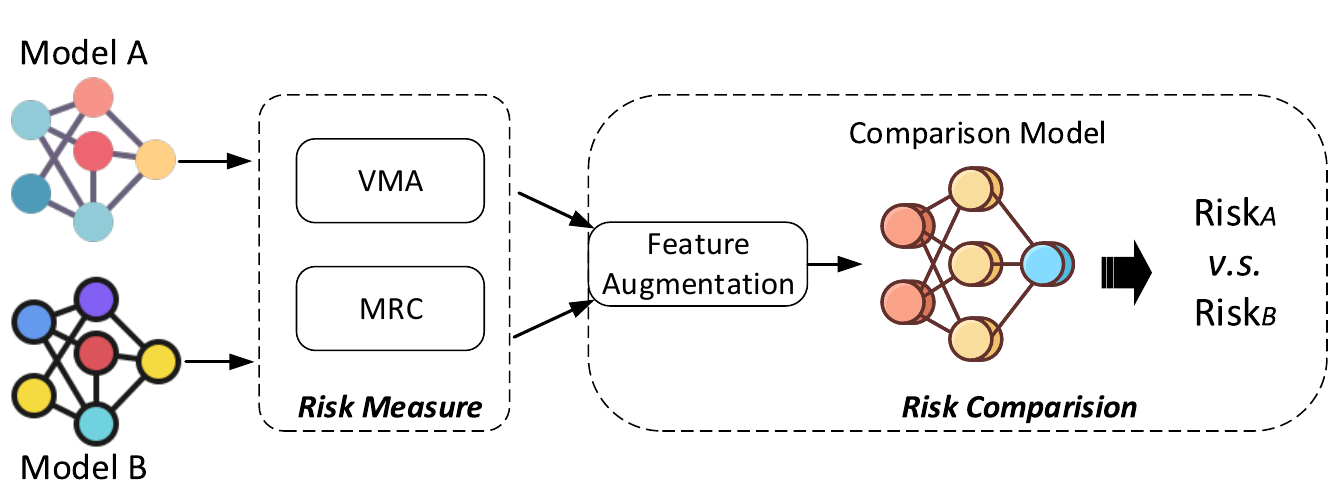}
	\caption{The framework of MER-Inspector.}
	\label{ML_Insepctor_framework}
\end{figure}

Since these two metrics complement each other, we can exploit both to compare the relative risks of two models. Figure \ref{ML_Insepctor_framework} shows MER-Inspector, the model risk comparison framework of models $A$ and $B$. It consists of two phases, i.e., risk measure and risk comparison, and outputs the relation between $\text{Risk}_A$ and $\text{Risk}_B$.

%
\begin{itemize}
\item Risk measure. First, the evaluation metric vectors (VMA and MRC) $\mathbf{r}_A$ and $\mathbf{r}_B$ for models $A$ and $B$ are obtained. VMA is measured on the testing dataset, and MRC is calculated according to Algorithm \ref{Alg1}.

\item Risk comparison. We can train a comparison model, e.g., a fully connected neural network as in Section~\ref{experiment}, to output bit 0 or 1 ($\text{Risk}_A$ exceeds $\text{Risk}_B$ or vice versa). Since there are now only two dimensions in the risk vector, we adopt Feature Augmentation (FA) to expand the vector from $\{\mathbf{r}_A,\mathbf{r}_B\}$ to $\{\mathbf{r}_A,\mathbf{r}_B,\mathbf{r}_A-\mathbf{r}_B\}$ as the input of comparison model. 
\end{itemize}

\section{Experiments}
\label{experiment}
\subsection{Experiment Setup}
\begin{figure}[t]
    \centering
    \subfigure[CIFAR-10, MRC]{
        \includegraphics[width=0.22\textwidth]{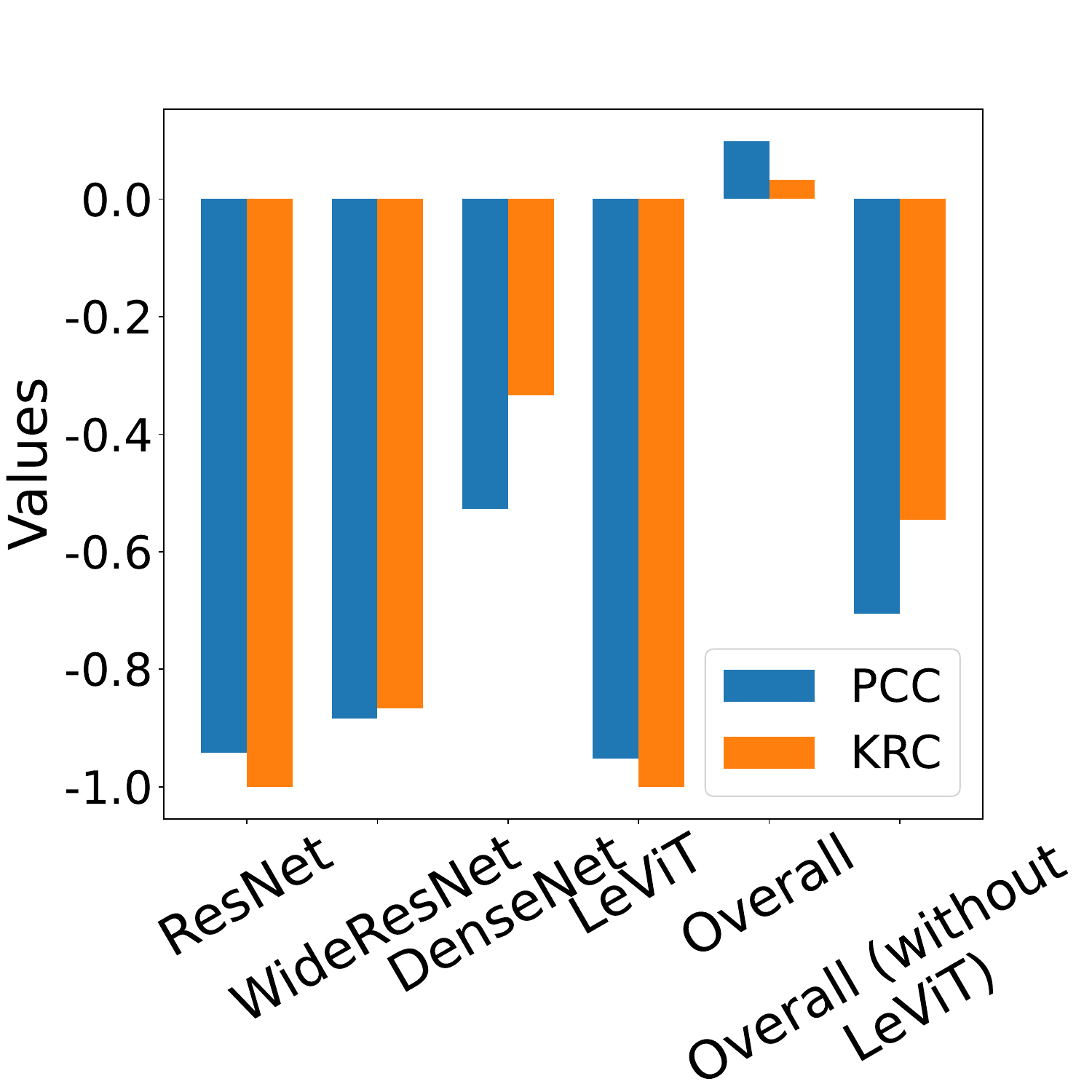}
        \label{fig:rc_pcc_cifar10}
    }
    \subfigure[STL-10, MRC]{
        \includegraphics[width=0.22\textwidth]{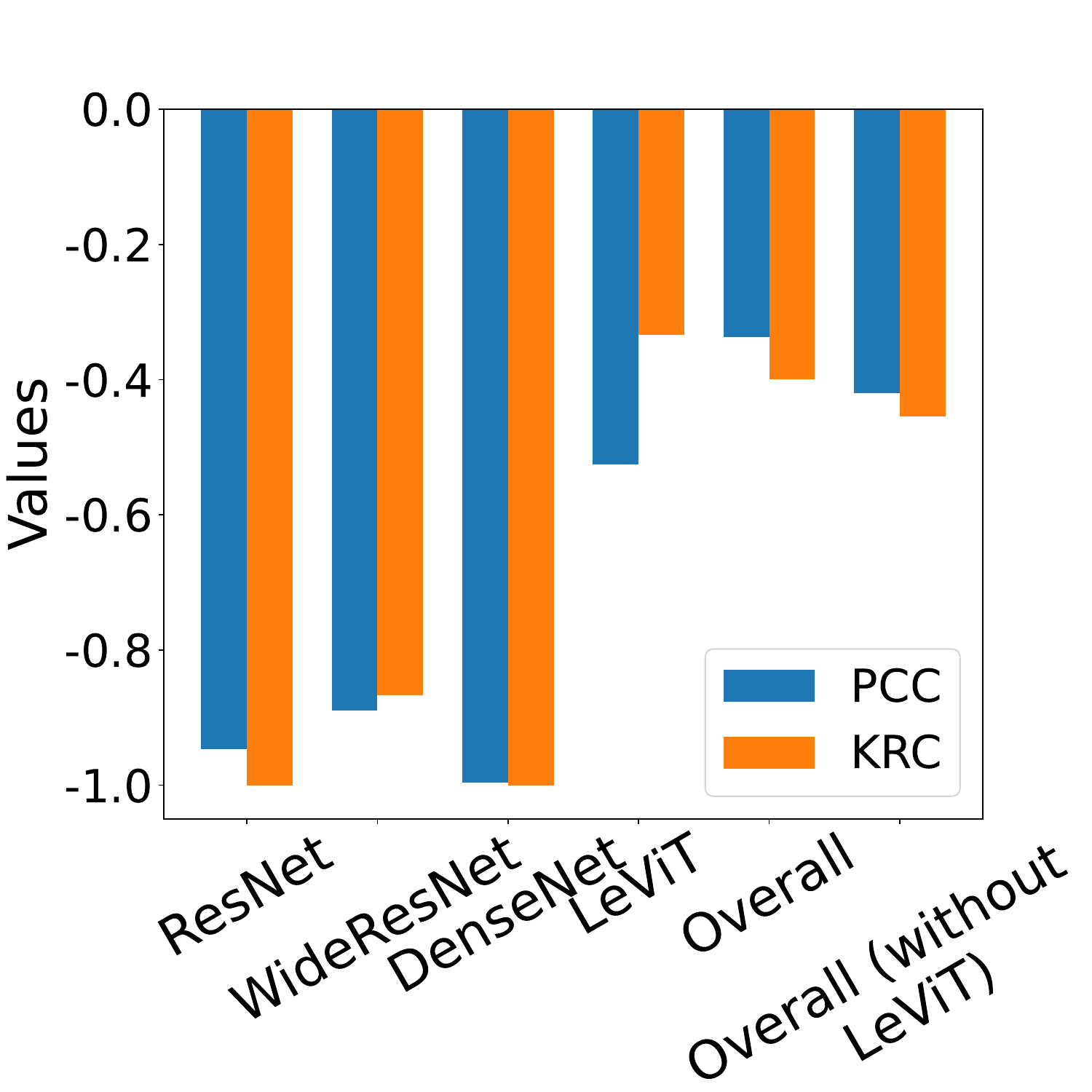}
        \label{fig:rc_pcc_stl10}
    }
    \subfigure[CIFAR-10, VMA]{
        \includegraphics[width=0.22\textwidth]{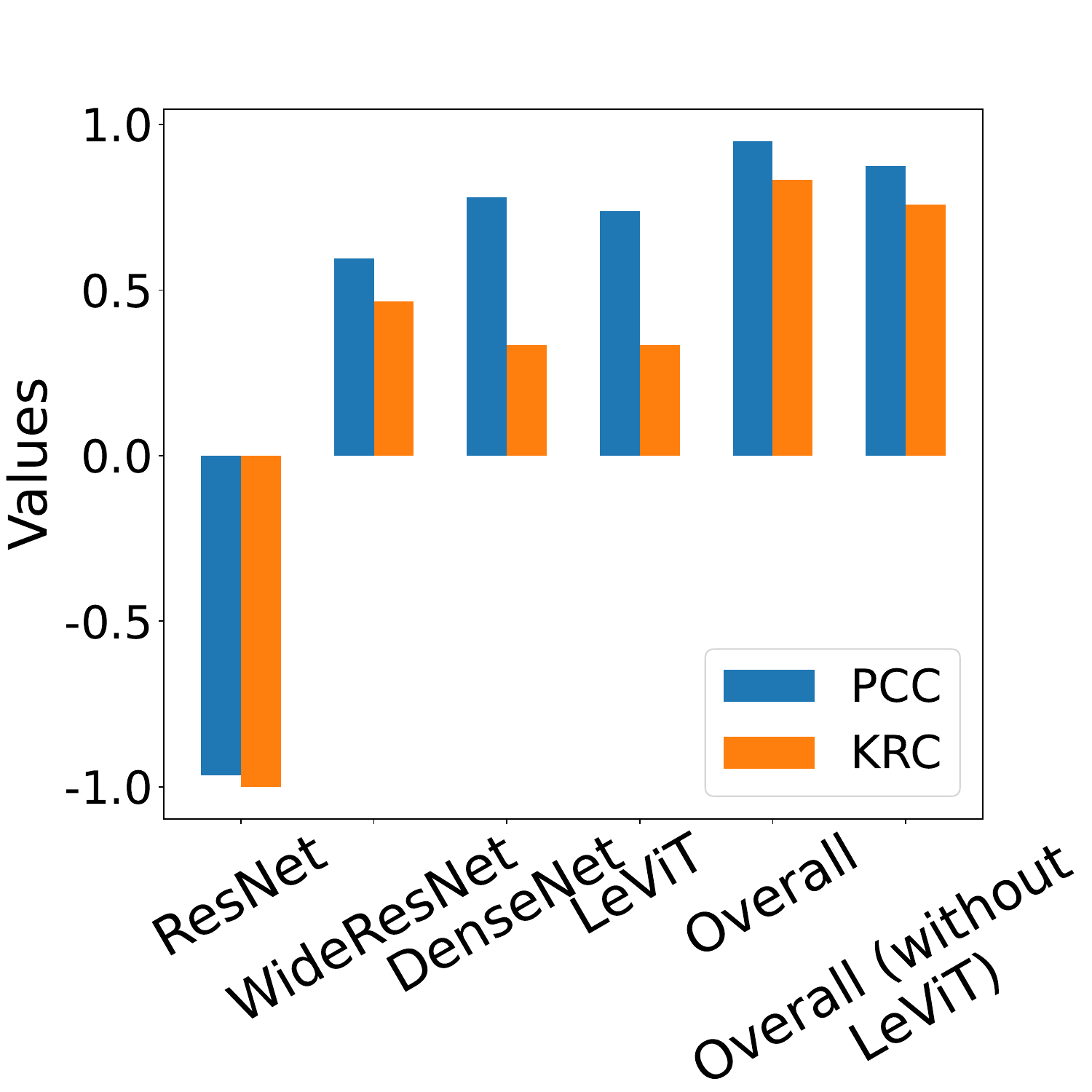}
        \label{fig:vma_pcc_cifar10}
    }
    \subfigure[STL-10, VMA]{
        \includegraphics[width=0.22\textwidth]{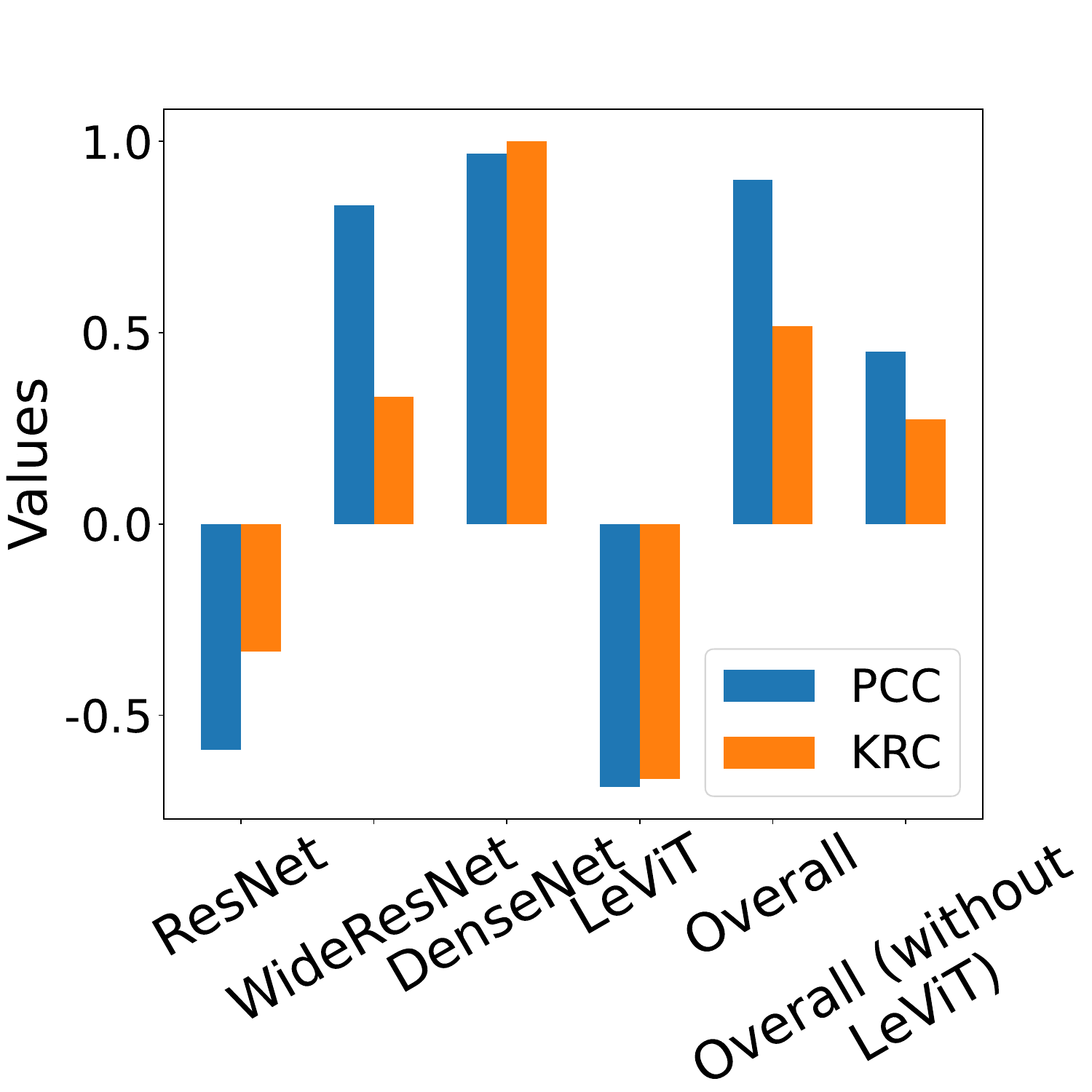}
        \label{fig:vma_pcc_stl10}
    }
    \caption{\textbf{Left:} PCC and KRC performance between MRC and attack fidelity on (a) CIFAR-10 and (b) STL-10. \textbf{Right:} PCC and KRC performance between VMA and attack fidelity on (c) CIFAR-10 and (d) STL-10.}
    \label{fig:combined_figures}
    \vskip -0.1in
\end{figure}

\textbf{Datasets \& Victim Models.} We use 5 evaluation datasets: CIFAR-10 \cite{Krizhevsky09learningmultiple}, CIFAR-100 \cite{Krizhevsky09learningmultiple}, FashionMNIST \cite{xiao2017fashion}, STL-10 \cite{coates2011analysis}, and CelebA \cite{liu2015faceattributes}. For each dataset, we train 16 victim models across four architecture groups for 200 epochs using the SGD optimizer with a learning rate of 0.01 (400 epochs for LeViT models) to minimize the softmax Cross-Entropy(CE) loss. The four architecture groups are ResNet \cite{he2016deep} (ResNet22, ResNet32, ResNet44), WideResNet \cite{zagoruyko2016wide} (WideResNet22-2, WideResNet22-4, WideResNet22-8, WideResNet28-2, WideResNet34-2.WideResNet40-2), Densenet \cite{huang2017densely} (Densenet121, Densenet169, Denset201) ), and LeViT \cite{graham2021levit} (LeViT-128, LeViT-192, LeViT-256, LeViT-384). The VMA difference is relatively small between models within the same group and large between different groups. The detailed
datasets and setup are introduced in Appendix~\ref{Datasets_Models} and Appendix~\ref{Detailed_Experimental_Setup}.

\textbf{Model Extraction Setting.} In the MEA, the attacker uses the same architecture and hyper-parameters as the victim model, and uses the training dataset of the victim model for querying. Each surrogate model is trained for 100 epochs using CE loss and SGD optimizer, and the maximum attack fidelity on the testing dataset during training is considered as the ground truth for assessing the risk.

\textbf{MER-Inspector Implementation.}\footnote{Our code is avaidable at \url{https://github.com/XinweiZhang1998/MER_Inspector}.}
The architecture of the comparison model is a fully connected neural network with three hidden layers containing 64, 64, and 32 neurons, respectively. The activation function for hidden layers is ReLU and is sigmoid for the output layer. The model is trained for 500 epochs using the Adam optimizer with a learning rate of 1e-4 to minimize the binary CE loss.
We select 16 model architectures and 5 datasets to verify the performance of MER-Inspector. In each dataset, two models are randomly paired. Their risk metrics are used as input to the comparison model, with label "0" meaning the difference in attack fidelity between the two models is positive, and label "1" otherwise. We then switch the order of the two models and get a total of 1,200 sets of data, a.k.a., \textit{all dataset}. Among them, 270 sets are from pairs within the same group, a.k.a., the \textit{intra-group dataset}. The rest 930 sets are from pairs in different groups, a.k.a., the \textit{inter-group dataset}. 80\% of each dataset is used for training and the rest 20\% for testing.

\begin{table}[t]
	\caption{Comparative accuracy of MER-Inspector.}
	\label{table:ML_Inspector}
	\begin{center}
		\begin{small}
				\begin{tabular}{p{2cm}ccc}
					\toprule
					Dataset & VMA & MRC & VMA+MRC \\
					\midrule
					Intra-Group  & 53.70\% & \textbf{88.89\%} & 79.63\% \\
					Inter-Group & 89.78\% & 55.38\% & \textbf{93.01\%}\\
					All   &  85.00\% & 70.42\% & \textbf{89.58\%} \\
					All (w/o FA)   &  83.33\% & 67.92\% & \textbf{86.25\%} \\
					\bottomrule
				\end{tabular}
		\end{small}
	\end{center}
\end{table}

\textbf{Evaluation Metric \& Baseline.} We employ PCC \cite{cohen2009pearson} and KRC \cite{abdi2007kendall} to evaluate how well the proposed metrics are aligned with the risk. Both metrics range from -1 to 1, where values closer to 1 indicate a strong positive correlation, values near -1 indicate a strong negative correlation, and values around 0 suggest no correlation. Furthermore, we use Comparative Accuracy (CAcc) to evaluate the effectiveness of the MER-Inspector. It is the ratio of pairs whose risks are successfully compared to the total compared pairs. The VMA \cite{zhang2023SecurityNet} alone serves as the baseline.

\subsection{Results}
\label{results}
\textbf{Effectiveness of Metrics.} First, we verify the effectiveness of MRC and VMA. We compute the MRC in the default settings when $L$ is 400, $\eta$ is 0.5, and $q$ is 0.5. Figure \ref{fig:rc_pcc_cifar10} and \ref{fig:rc_pcc_stl10} show the PCC and KRC performance between MRC and attack fidelity under different model architecture groups and datasets. The \textit{overall} considers all 16 architectures. The results show that there is a strong negative correlation between MRC and attack fidelity within each model structure group. However, the correlation on CIFAR10 is weak positive. This is due to the large differences between the transformer-based model (LeViT) and other models, which makes the NTK theory difficult to capture this difference. The other metric, VMA, can compensate for this shortcoming. From Figures \ref{fig:vma_pcc_cifar10} and \ref{fig:vma_pcc_stl10}, we find that overall there is a positive correlation between VMA and attack fidelity, but when it comes to models in the same group, such correlation diminishes or even reverses, due to the small VMA gaps between models in the same group. This justifies our motivation to combine both metrics to compare risks. The results on other datasets are given in Appendix \ref{Effect_Metircs_Datasets}. 
\begin{figure}
\centering
	\includegraphics[width=0.9\columnwidth]{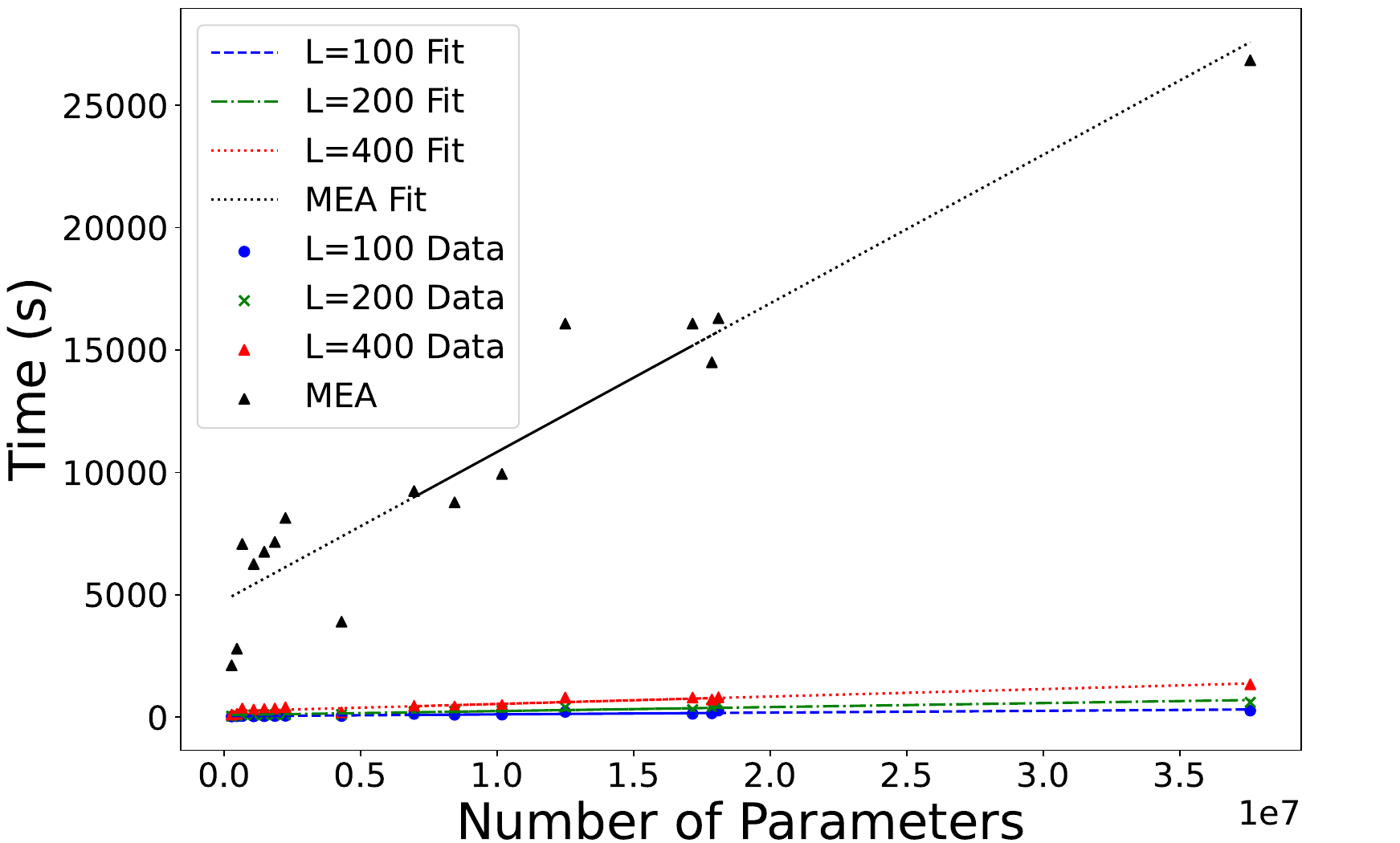}
	\caption{Time cost for calculating MRC with varying sample numbers ($L$) on CIFAR-10.}
	\label{Time_Para}
\end{figure}

\textbf{Effectiveness of MER-Inspector.}
Table \ref{table:ML_Inspector} compares CAccs across various datasets using different metrics. The results indicate that MRC effectively evaluates risks within similar architecture models (intra-group) with a CAcc of 88.89\%. Meanwhile, VMA performs better in identifying risks between models (inter-group) with significant accuracy differences, achieving a CAcc of 89.78\%.
Combining these two metrics can significantly improve the overall accuracy (CAcc of 89.58\%) of risk comparison on any two models. Furthermore, by comparing the last two columns of Table \ref{table:ML_Inspector}, we observe an improvement in CAcc when FA is utilized.

\textbf{Cost Analysis.} Figure~\ref{Time_Para} illustrates that the time cost increases with the number of parameters and the number of selected samples. We adopt the simplest MEA, where the attacker queries all samples and trains for 100 epochs. The cost will significantly increase if a more complex MEA is used (e.g., selecting the most uncertain samples for querying and training each iteration). Our results show that the time cost of MRC calculation is significantly lower than that of performing even a simple MEA. When we select 100 samples to calculate the MRC, the clock time is only 269 seconds for a model with nearly 40 million parameters, significantly less than the time cost of 2,980 seconds to execute an MEA for risk assessment, which requires full training of the surrogate model.

\begin{table}
	\caption{RCC and KRC performance between MRC and attack fidelity in the LeViT group on CIFAR-10.}
	\label{table:L}
	\begin{center}
		\begin{small}
				\begin{tabular}{lcccr}
					\toprule
					Metric & $L=100$ & $L=200$ & $L=400$ \\
					\midrule
					PCC  & -0.7508 & -0.8463 & -0.9523 \\
					KRC  &  -0.3333 & -0.6667 & -1.0 \\
					\bottomrule
				\end{tabular}
		\end{small}
	\end{center}
    	\vskip -0.1in
\end{table}


\begin{table*}
	\caption{RCC / KRC performance between MRC and attack fidelity on CIFAR-10 versus different $\eta$. A large $\eta$ means that more hard samples (samples with the most significant margin) are selected.}
	\label{table:DR}
	\begin{center}
		\begin{small}
					\begin{tabular}{lccccc}
						\toprule
						Group& $\eta=0$ & $\eta=0.25$ & $\eta=0.5$ & $\eta=0.75$ & $\eta=1$ \\
						\midrule
						ResNet  & 0.199/0.333 & -0.356/-0.333 &  -0.942/-1.0 & -0.974/-1.0 & \textbf{-0.995/-1.0}\\
						WideResNet  & \textbf{-0.946/-1.0} & -0.913/-0.867 & -0.884/-0.867& -0.888/-0.867& -0.912/-0.867\\
						DenseNet  &  \textbf{-0.950/-1.0}&-0.998/-1.0 & -0.528/-0.333& -0.944/-0.333&-0.753/-0.333\\
						LeViT  & -0.375/-0.333& -0.717/-0.333 &-0.952/-1.0 & \textbf{-0.993/-1.0}& -0.836/-0.333\\
						\bottomrule
				\end{tabular}
		\end{small}
	\end{center}
\end{table*}

\begin{table}
	\caption{PCC / KRC performance versus different $q$.}
	\label{table:DQ}
	\begin{center}
		\begin{small}
				\begin{tabular}{lccc}
					\toprule
					Group& $q=0.05$ & $q=0.5$  & $q=5$ \\
					\midrule
					\multirow{1}{*}{\shortstack{ResNet}}  & -0.779/-0.333&-0.942/-1.0&-0.324/-0.333 \\
					\multirow{1}{*}{\shortstack{WideResNet}}
					& -0.719/-0.733 & -0.884/-0.867& -0.591/-0.414\\
					\multirow{1}{*}{\shortstack{DenseNet}}
					&-0.587/-0.333& -0.528/-0.333& -0.526/-0.333\\
					\multirow{1}{*}{\shortstack{LeViT}}
					& -0.860/-0.667&-0.952/-1.0 & -0.953/-1.0 \\
					\bottomrule
				\end{tabular}
		\end{small}
	\end{center}
\end{table}

\begin{table}
\caption{Comparison of standard training and adversarial training.}
\label{Adv_train}
\begin{center}
\begin{small}
\resizebox{0.48\textwidth}{!}{
\begin{tabular}{c|ccc|ccc}
\toprule
\multirow{2}{*}{Network} & \multicolumn{3}{c|}{Standard-training} & \multicolumn{3}{c}{Adv-training} \\
& VMA (\%) & Attack Fidelity (\%) & MRC & VMA (\%) & Attack Fidelity (\%) & MRC \\
\midrule
ResNet20 & 90.94 & 81.35 & 0.00353 & 77.59 & 77.37 & 0.01418 \\

ResNet32 & 91.66 & 80.88 & 0.00405 & 76.30 & 73.66 &  0.01500\\

ResNet44 & 92.41 & 80.31 &  0.00736& 82.29 & 77.69 &  0.00994\\
\bottomrule
\end{tabular}}
\end{small}
\end{center}
\label{table:training_comparison}
\end{table}

\subsection{Ablation Study}
\textbf{Impact of Number of Selected Samples $L$.} In the previous experiments, $L$ is set to 400 and only a small part of the training data is used. For example, only 0.8\% of the training samples are selected on CIFAR10. A larger $L$ leads to a more accurate approximation of MRC.
We select the LeViT group as an example to illustrate this point. 
As shown in Table \ref{table:L}, the PCC and RKC are closer to -1 as $L$ increases. The detailed results of the other models and other datasets are in Appendix \ref{Detailed_results}. 
Besides, models with fewer parameters or simpler architectures, such as ResNet, can be accurately measured even when $L=100$. Therefore, a trade-off needs to be made between computational complexity and accuracy.

\textbf{Impact of Difficulty Ratio $\eta$.}
Table \ref{table:DR} compares the performance of different groups under different $\eta$.
The results show that simple samples effectively indicate risks associated with well-generalized models (such as WideResNet and DenseNet), while hard samples are important for models with poorer generalization (such as ResNet and LeViT). This is consistent with \citet{sorscher2022Neural} where models with poor generalization are more likely to learn hard samples close to the boundary. To fairly compare models from different model groups, setting $\eta$ to 0.5 seems appropriate in our experiments.

\textbf{Impact of Threshold $q$.}
To maintain the fine-grained information of the NTK matrix, $q$ should be set as small as possible while ensuring the positive definiteness of the NTK matrix of most models.
Table \ref{table:DQ} compares the performance under different $q$, and we find that the performance is poor when setting $q=0.05/5$. This is because only 18.75\% of the models exhibit positive definiteness in their NTK matrices when $q=0.05$, causing inaccurate results. When $q=5$, although 93.75\% matrices are positive definite, a lot of fine-grained information is lost. In our experiments, we set $q$ to 0.5, which ensures that the NTK matrices of most (87.5\%) models remain positive definite and retain a large amount of fine-grained information.

\section{Discussion}
\label{discussion}
\subsection{Practical Threat Model}
\label{practical_tm}

\begin{table}[h]
	\centering
	\caption{Performance of MER-Inspector under different attackers' capabilities.  }
	\begin{small} 
		\resizebox{0.48\textwidth}{!}{
		\begin{tabular}{c|ccc|ccc}
			\toprule
			\multirow{2}{*}{\diagbox{Dataset}{Case}}  & \multicolumn{3}{c|}{CIFAR100 $\longrightarrow$ CIFAR10} & \multicolumn{3}{c}{ Only Label} \\ 
			& VMA & MRC & VMA+MRC & VMA & MRC & VMA+MRC \\ \midrule
			All & 88.33\% & 55.00\% & \textbf{90.42\%} & 91.67\% & 51.67\% & \textbf{92.08\%} \\ 
			Intra-Group & 55.56\% & \textbf{75.93\%} & 72.22\% & 72.22\% & 68.52\% & \textbf{79.63\%} \\ 
			Inter-Group & 97.31\% & 48.39\% & \textbf{97.85\%} & \textbf{97.31\%} & 47.31\% & 96.77\% \\ 
			\bottomrule
		\end{tabular}}
	\end{small}
	\label{practical_scenario}
\end{table}

\begin{table}[!t]
	\centering
	\caption{Performance of MER-Inspector under different attack strategies. }
	\begin{small} 
		\resizebox{0.48\textwidth}{!}{
		\begin{tabular}{c|ccc|ccc}
			\toprule
			\multirow{2}{*}{\diagbox{Dataset}{Case}} & \multicolumn{3}{c|}{Uncertain} & \multicolumn{3}{c}{K-center} \\ 
			& VMA & MRC & VMA+MRC & VMA & MRC & VMA+MRC \\ \midrule
			All & \textbf{91.25\%} & 52.50\% & \textbf{91.25\%} & 90.00\% & 52.50\% & \textbf{91.25\%} \\ 
			Intra-Group & 64.81\% & \textbf{72.22}\% & \textbf{72.22}\% & 57.41\% & \textbf{72.22\%} & \textbf{72.22\%} \\ 
			Inter-Group & 99.46\% & 47.31\% & \textbf{100.00\%} & 99.46\% & 47.31\% & \textbf{100.00\%} \\ 
			\bottomrule
		\end{tabular}}
	\end{small}
	\label{different_strategy}
\end{table}

Although our assumption of the most powerful attacker is made to analyze the maximum possible risk, our proposed MRC metric is based on model complexity, can also align with model extraction risk in more realistic scenarios (e.g., when the attacker can only access model labels, or does not know the training samples, or employs different attack strategies). As shown in Table \ref{practical_scenario}, we verify the performance under two cases: 1. the attacker uses the surrogate dataset (e.g., CIFAR100) to steal the victim model (trained by CIFRA10) rather than using the victim model's training dataset CIFAR10; 2. the attacker only accesses the victim model's output label rather than output probabilities. We use the trained comparison model to compare the model risk under these two practical cases, and our results show that our metrics are still effective. We also evaluate the performance when the attacker is constrained by a query budget of 20,000 and employs two distinct query strategies as described in ActiveThief [11]: (1) the uncertainty-based strategy and (2) the K-center strategy. Using a trained comparison model, we assess the extraction risk under both strategies and the results in Table \ref{different_strategy} demonstrate that our proposed metrics remain effective.

\subsection{Understanding Current Attacks and Defenses}
By utilizing the theoretical metric known as MRC, we can deepen our understanding of existing attack and defense strategies.

\textbf{Attacks.} Current attacks aim to enhance query efficiency through active learning by identifying both informative (hard) and representative (simple) samples \cite{orekondy2019Knockoff, yu2020CloudLeak, AL_ME}. MRC provides us with an opportunity to gain fresh insights into the rationale behind utilizing these samples in queries. 
In Appendix \ref{detailed_eat}, we compare the MRC scores calculated for randomly selected samples, as well as hard and simple samples chosen based on difficulty ratio $\eta$ when $L=400$ on CIFAR-10.
The results show that MRC scores derived from hard and easy samples yield lower MRC values compared to scores calculated using randomly selected samples.
In specific, when the weight change of the victim model is projected onto the span $\{\nabla_{\theta_v} \mathbf{f}_{\theta_{ 0 }}(\mathbf{x}) \}$ consisting of the samples with the largest and smallest margins shows smaller differences from the original weight change. Thus the surrogate model's weight change $\mathbf{P}_{\theta_v} \bigtriangleup_{\theta_v}$ is closed to the victim model's weight change $\bigtriangleup_{\theta_v}$. In this way, we provide a theoretical explanation for why selecting useful samples is crucial for MEAs.


\textbf{Defences.} We can also use MRC to understand the defense methods on the model itself, such as adversarial training \cite{jiang2023comprehensive}. We use Project Gradient Descent (PGD) \cite{madry2018towards} to generate adversarial examples. 
The perturbation magnitude of the input \( x \) is constrained to $\epsilon = 0.1 $. The step size $\alpha$ is set to 0.01 during the generation of adversarial samples. The PGD process of the optimized perturbation is performed three times. Then, we use these generated adversarial examples to train 200 epochs as the victim model. The other setups are the same as in Appendix \ref{Detailed_Experimental_Setup}. Table \ref{Adv_train} compares the results of standard training and adversarial training on CIFAR-10. As expected, adversarial training leads to an increase in MRC scores and a decrease in attack fidelity. In addition to adversarial training, we can utilize MRC to validate additional model defenses and investigate novel defense strategies that inherently mitigate the model extraction risk by enhancing the MRC score.

\section{Conclusion}
\label{conclusion}
This paper provides a theoretical analysis of MEAs from an attack-agnostic perspective and proposes analytical metrics to evaluate model extraction risks for the first time. We start by theoretically analyzing MEAs using NTK theory and provide two bounds on attack performance. Next, we introduce two metrics, MRC and VMA, which jointly align with the model extraction risk. Using these metrics, we present the MER-Inspector framework that can compare risks on any two models. Experimental results verify the effectiveness of these metrics and MER-Inspector. 
This work aids in understanding and validating current model extraction attacks and defenses, while also inspiring future exploration of new strategies. As for future work, we will continue to explore additional metrics and new attack and defense strategies. 

\begin{acks}
This work was supported by the National Natural Science Foundation of China (Grant No: 92270123 and 62372122), and the Research Grants Council, Hong Kong SAR, China (Grant No:  15226221, 15209922, 15208923, 15210023, and 15224124). 
\end{acks}

\bibliographystyle{ACM-Reference-Format}
\balance
\bibliography{My_ref}

\appendix

\appendix
\section{MRC Approximation Algorithm}
\label{A1}
The MRC approximation algorithm is detailed in Algorithm \ref{Alg1}.
\begin{algorithm}[t!]
	\caption{MRC Approximation Algorithm}
	\begin{algorithmic}[1]
		\label{Alg1}
		\REQUIRE Victim model $M$, training dataset $\mathbb{D}$, threshold $q$, the number of selected samples $L$, difficulty ratio $\eta$
		\ENSURE MRC $r_{rc}$
		\STATE Select $L$ samples as dataset $\mathbb{D}_S$ according to difficulty ratio $\eta$ from $\mathbb{D}$
		\FOR{each data point $x_i$ in $\mathbb{D}_S$}
		\STATE Compute gradients of $M$ on $x_i$
		\ENDFOR
		\STATE Calculate NTK matrix using their gradients according to Eq. (\ref{NTK_computer})
		\STATE Compute eigenvalues and eigenvectors of NTK matrix
		\FOR{each eigenvalue $\lambda_i$}
		\IF{$\lambda_i < q$}
		\STATE $\lambda_i \gets q$
		\ENDIF
		\ENDFOR
		\STATE Reconstruct NTK matrix from adjusted eigenvalues and eigenvectors
		\STATE Computer the output probability change of $\mathbb{D}_S$
		\STATE Computer the MRC using Eq. (\ref{r1})
	\end{algorithmic}
\end{algorithm}

\section{Proof}
\label{proof}
\subsection{Basic Theory}
\label{rademacher_complexity}
Rademacher complexity quantifies the expressive power of a machine-learning model and the inherent risk of overfitting. The definition is given as follows \cite{mohri2018foundations}.

\begin{definition} [Rademacher complexity]
	Given samples \(S = \{x_1, x_2, \ldots, x_n\}\) of size \(n\) drawn i.i.d. from some distributions, and a class of functions \(F\), the empirical Rademacher complexity of \(F\) with respect to \(S\) is defined as:
	\begin{equation}
		\hat{\mathfrak{R}}_n(F) = \frac{1}{n} \mathbb{E}_{\sigma} \left[ \sup_{f \in F} \sum_{i=1}^{n} \sigma_i f(x_i) \right]
	\end{equation}
	where \(\sigma = (\sigma_1, \sigma_2, \ldots, \sigma_n)\) is a vector of independent Rademacher random variables (i.e., each \(\sigma_i\) takes values +1 or -1 with equal probability).
	The (population) Rademacher complexity of \(F\) is then defined as the expectation of the empirical Rademacher complexity over all possible samples of size \(n\), which is given by:
	\begin{equation}
		\mathfrak{R}_n(F) = \mathbb{E}_{S}[\hat{R}_n(F)].
	\end{equation}
\end{definition}

\textit{Remarks.} Rademacher complexity measures the complexity of the hypothesis space $\mathcal{H}$ in which the model resides and is a powerful indicator of the model extraction risk. If the complexity of the hypothesis space is higher, the behavior of the model will be more difficult to predict and understand, which may make model extraction attacks more difficult. However, this complexity cannot be used directly to measure risk due to the high computational cost and separation with models. Firstly, the computation of Rademacher complexity often involves optimizing over a large hypothesis space, which can be computationally intensive, especially for complex models or large datasets. Furthermore, we rely on empirical estimates of Rademacher complexity to obtain accurate estimates, which requires repeated estimation of the complexity multiple times, which further increases the computational burden. Secondly, Rademacher complexity is based on the hypothesis space $\mathcal{H}$ and the nature of the data rather than on a specific model. Therefore, it may not provide in-depth insights into the behavior and performance of a specific model, which may lead to an inaccurate measurement of model complexity.

For a multi-label classification task, given a labeled example $(\mathbf{x},y)$, the prediction margin is defined as $\mathcal{M}(\mathbf{x},y)=f^y(\mathbf{x})-\max_{y'\ne y} f^{y'}(\mathbf{x}) $. Considering the class of functions $\mathcal{F} =\{ f \in \mathcal{H}: \parallel f \parallel \leq M\}$ for some $M>0$, we can give the generalization bound based on the Rademacher complexity in the RKHS $\mathcal{H}$ by the theorem proposed in \cite{kuznetsov2015rademacher}.



\begin{theorem}[Generalization bound]
Considering that the training dataset \(\mathbb{D}\) comprises \(N\) i.i.d. is\
samples, each associated with one of \(K\) labels, derived from the input distribution \(P_{\mathbf{x}}\). Assume $\kappa  = \sup_{\mathbf{x} \in P_\mathbf{x}} k(\mathbf{x},\mathbf{x}) < \infty $. Fix any constant $M > 0$ and $\gamma > 0$. Let $M_{0}=\lceil\gamma \sqrt{N} /(4 K \sqrt{\kappa})\rceil$. Then with probability at least $1-\delta $, every function $f \in \mathcal{H}$  has the following bound:  
	\begin{equation}
        \label{ref:generalization_bound}
		\begin{aligned}
			\mathbb{P}_{X, Y}\left(\mathcal{M}(\mathbf{x},y) \leq 0\right) \leq & \frac{1}{N} \sum_{i=1}^{N} \mathbf{1}\left\{\mathcal{M}(\mathbf{x}^{(i)},y^{(i)}) \leq \gamma\right\} 
			\\
            &+\frac{4 K}{\gamma} \hat{\mathfrak{R}}_n(F)
			+3 \sqrt{\frac{\log \left(2 M_{0} / \delta\right)}{2 N}}.
		\end{aligned}
	\end{equation}
\end{theorem}
Based on this generalization bound, we further give two theoretical bounds on the attack performance.

\subsection{Proof of Theorem \ref{fidelity_gap_bound}}
\label{proof4.3}
\begin{theorem}[Theorm \ref{fidelity_gap_bound} restated]
	 Assume that \(\kappa = \sup_{\mathbf{x} \sim P_{\mathbf{x}}} k(\mathbf{x},\mathbf{x}) < \infty\). Define a constant \(\gamma > 0\), and let \(M_{0} = \left\lceil\frac{\gamma \sqrt{N}}{4 K \sqrt{\kappa}}\right\rceil\). Subsequently, with a probability of at least \(1 - \delta\), for every function \(\bigtriangleup_{\mathbf{f}_{\theta_s}(\mathbf{x})} \in \mathcal{H}\), the fidelity gap on $N$ samples can be bounded:
	 \begin{equation}
	 	\begin{aligned}
	 		\mathcal{G}_N = & ~\underset{\mathbf{x} \sim P_{\mathbf{x}}}{\mathbb{P}} [ \mathcal{M}(\mathbf{x},y_v,y_s) \leq 0 ] \\ \leq
	 		& \frac{1}{N} \sum_{i=1}^{N} \mathbf{1}\left\{\mathcal{M}(\mathbf{x}^{(i)},y_v^{(i)},y_s^{(i)}) \leq \gamma\right\}\\
	 		&+\frac{4 K(\bigtriangleup^{\top}_{{\mathbf{f}}_{\theta_v}(\mathbf{x})}\Theta^{-1}\bigtriangleup_{{\mathbf{f}}_{\theta_v}(\mathbf{x})})}{\gamma N} \sqrt{\text{Tr} (\Theta)}\\
	 		&+3 \sqrt{\frac{\log \left(2 M_{0} / \delta\right)}{2 N}},
	 	\end{aligned}
	 \end{equation}
	 where $\text{Tr} (\Theta)$ represents the trace of the victim model's NTK matrix $\Theta$.
\end{theorem}
\begin{proof}
    \citet{harutyunyan2023Supervision} give the upper bound of the empirical Rademacher complexity of $\mathcal{F}$, i.e., 
    \begin{equation}
        \hat{\mathfrak{R}}_n(F) \leq \frac{M}{N} \sqrt{\text{Tr} (\Theta)}.
    \end{equation}
    Under the assumptions of the NTK theory and no two samples are the same, the NTK matrix is the positive definite matrix, and $\Theta$ is full rank surely. Based on the Eq. (\ref{F_change}), the solution is 
    \begin{equation}
	\bigtriangleup_{\mathbf{f}_{\theta_s}(\mathbf{x})} =\sum_{i=1}^{N}\alpha_i k(\mathbf{x},\mathbf{x}^{(i)})= \Theta \boldsymbol{\alpha}.
    \end{equation}

    When $\bigtriangleup_{{f}_{\theta_s}(\mathbf{x})}$ in $\mathcal{H}$ has zero empirical loss,  we have $\bigtriangleup_{\mathbf{f}_{\theta_s}(\mathbf{x})} = \bigtriangleup_{\mathbf{f}_{\theta_v}(\mathbf{x})}$, and 
    the norm of $\bigtriangleup_{\mathbf{f}_{\theta_s}(\mathbf{x})}$ can be expressed as
    \begin{equation}
    \label{fh_change}
    	\|\bigtriangleup_{\mathbf{f}_{\theta_s}(\mathbf{x})}\|_{\mathcal{H}}^{2}=\boldsymbol{\alpha}^{\top} \Theta \boldsymbol{\alpha} = \bigtriangleup^{\top}_{{\mathbf{f}}_{\theta_v}(\mathbf{x})}\Theta^{-1}\bigtriangleup_{{\mathbf{f}}_{\theta_v}(\mathbf{x})}.
    \end{equation}
    Therefore, any optimal solution to the optimization problem in Section \ref{problem_setup} has a norm at most $\bigtriangleup^{\top}_{{\mathbf{f}}_{\theta_v}(\mathbf{x})}\Theta^{-1}\bigtriangleup_{{\mathbf{f}}_{\theta_v}(\mathbf{x})}$.
    Considering the class of functions $\mathcal{F} =\{ \bigtriangleup_{{\mathbf{f}}_{\theta_s}(\mathbf{x})} \in \mathcal{H}: \parallel \bigtriangleup_{{\mathbf{f}}_{\theta_s}(\mathbf{x})} \parallel \leq M\}$ for some $M>0$, we can get that
    \begin{equation}
        M \leq \bigtriangleup^{\top}_{{\mathbf{f}}_{\theta_v}(\mathbf{x})}\Theta^{-1}\bigtriangleup_{{\mathbf{f}}_{\theta_v}(\mathbf{x})}.
    \end{equation}
    and 
    \begin{equation}
    \label{rf}
        \hat{\mathfrak{R}}_n(F) \leq \frac{\bigtriangleup^{\top}_{{\mathbf{f}}_{\theta_v}(\mathbf{x})}\Theta^{-1}\bigtriangleup_{{\mathbf{f}}_{\theta_v}(\mathbf{x})}}{N} \sqrt{\text{Tr} (\Theta)}.
    \end{equation}
    After substituting Eq. (\ref{rf}) into Eq. (\ref{ref:generalization_bound}), we can prove this theorem.
\end{proof}

\subsection{Proof of Theorem \ref{Generalization_risk_bound}}
\label{proof4.4}

\begin{theorem}[Theorem \ref{Generalization_risk_bound} restated]
    Given the fidelity gap bound \( \mathcal{G}_N\) and the generalization error bound of the victim model \(\mathcal{R}_N^v\). Then, the generalization error of the surrogate model is bounded by,
	\begin{equation}
		\begin{aligned}
			\mathcal{R}_N^s &\leq \mathcal{G}_N + \mathcal{R}_N^v. \\
		\end{aligned}
	\end{equation}
\end{theorem}
\begin{proof}
Define the prediction margin $\mathcal{M}(\mathbf{x},y_v,y_s) = f_{\theta_s}^{y_v}(\mathbf{x}) - \max_{y'\ne {y_v}}f_{\theta_s}^{y'}(\mathbf{x})$, 
$\mathcal{M}(\mathbf{x},y_v,y) = f_{\theta_v}^{y}(\mathbf{x}) - \max_{y'\ne {y}}f_{\theta_v}^{y'}(\mathbf{x})$, 
$\mathcal{M}(\mathbf{x},y_s,y) = f_{\theta_s}^{y}(\mathbf{x}) - \max_{y'\ne {y}}f_{\theta_s}^{y'}(\mathbf{x})$, we have 
\begin{equation}
	\mathcal{G}_N =  \underset{\mathbf{x} \sim P_{\mathbf{x}}}{\mathbb{P}}\left[ \mathcal{M}(\mathbf{x},y_v,y_s) \le 0\right],
\end{equation}
and 
\begin{equation}
	\mathcal{R}^v_N =  \underset{\mathbf{x} \sim P_{\mathbf{x}}}{\mathbb{P}}\left[ \mathcal{M}(\mathbf{x},y_v,y) \le 0\right].
\end{equation}

Thus
\begin{equation}
    \begin{aligned}
    \mathcal{R}_N^s =&\underset{\mathbf{x} \sim P_{\mathbf{x}}}{\mathbb{P}}\left[ \mathcal{M}(\mathbf{x},y_s,y) \le 0\right].\\
    =& \underset{\mathbf{x} \sim P_{\mathbf{x}}}{\mathbb{P}}\left[ \mathcal{M}(\mathbf{x},y_s,y) \le 0 \wedge \mathcal{M}(\mathbf{x},y_v,y) \le 0 \right] \\
    &+ \underset{\mathbf{x} \sim P_{\mathbf{x}}}{\mathbb{P}}\left[ \mathcal{M}(\mathbf{x},y_s,y) \le 0 \wedge \mathcal{M}(\mathbf{x},y_v,y) > 0 \right] \\
    = & \underset{\mathbf{x} \sim P_{\mathbf{x}}}{\mathbb{P}}\left[ \mathcal{M}(\mathbf{x},y_s,y) \le 0 \wedge \mathcal{M}(\mathbf{x},y_v,y) \le 0 \right] \\
    &+ \underset{\mathbf{x} \sim P_{\mathbf{x}}}{\mathbb{P}}\left[\mathcal{M}(\mathbf{x},y_v,y_s) \leq 0 \right] \\
    \leq &  \underset{\mathbf{x} \sim P_{\mathbf{x}}}{\mathbb{P}}\left[ \mathcal{M}(\mathbf{x},y_v,y) \le 0\right] + \underset{\mathbf{x} \sim P_{\mathbf{x}}}{\mathbb{P}}\left[ \mathcal{M}(\mathbf{x},y_v,y_s) \le 0\right] \\
    = &  \mathcal{R}_N^v + \mathcal{G}_N.
    \end{aligned}
\end{equation}	

\end{proof}

\section{Datasets and Models}
\label{Datasets_Models}
We conduct experiments on five popular datasets:
\begin{itemize}
	\item CIFAR-10 \cite{Krizhevsky09learningmultiple} contains 60,000 32x32 color images across 10 different classes, such as automobiles, birds, and ships, with each class containing 6,000 images. The dataset is typically divided into 50,000 training images and 10,000 testing images.
	\item CIFAR-100 \cite{Krizhevsky09learningmultiple}  contains 60,000 32x32 color images across 100 different classes, with 600 images (500 for training, 100 for testing) per class.
	\item FashionMNIST \cite{xiao2017fashion} contains 70,000 28x28 grayscale images of fashion products from 10 categories, such as trousers, pullovers, and sandals. The dataset is typically divided into 60,000 training images and 10,000 testing images.
    \item STL-10 \cite{coates2011analysis} contains 5,000 96x96 color training images from 10 distinct classes, with each class represented in the testing set by 800 images.
    \item CelebA \cite{liu2015faceattributes} contains 202,599 facial color images, each associated with 40 binary attributes. Like \citet{liu2022MLDOCTOR}, we select and combine 3 attributes from 40 attributes, including HeavyMakeup, MouthSlightlyOpen, and Smiling, to form the labels of the target model, resulting in an 8-category classification. We resize the image pixels to 32x32 and randomly select 20\% of the entire dataset for training and testing, generating 32,554 training samples and 3,992 testing samples.
\end{itemize}

We conduct experiments on four famous architecture groups:
\begin{itemize}
    \item ResNet \cite{he2016deep} represents the classic approach of deep residual learning where skip connections facilitate the training of deeper networks. We use its variants ResNet20, ResNet32, and ResNet44.
    \item WideResNet \cite{zagoruyko2016wide} modifies the ResNet architecture by increasing width. We use its variants such as WideResNet22-2, WideResNet22-4, WideResNet22-8,  WideResNet28-2, Wide\\-ResNet34-2, and WideResNet40-2. In "WideResNetN-M", "N" denotes the depth of the network, while "M" signifies the multiplier for the number of convolutional kernels in comparison to the standard ResNet model.
    \item DenseNet \cite{huang2017densely} connects each layer to other layers in a feed-forward manner.  We use its variants such as DenseNet121, DenseNet169, and DenseNet201.
    \item LeViT \cite{graham2021levit} is a transformer-based architecture for image classification. We evaluate its various configurations, including LeViT-128, LeViT-192, LeViT-256, and LeViT-384.
\end{itemize}

\section{Experimental Setup of JBDA and MAZE}
\label{JDBA_MAZE_experiment}

\textbf{JBDA} \cite{papernot2017Practical} generates crafted adversarial examples for querying the victim model when the adversary has access to a few natural samples. Given a sample $x$ and its label $y$, a crafted query sample is constructed as
\begin{equation}
    x \leftarrow x - \lambda \nabla_x \mathcal{L}(y, f_s(x)),
\end{equation}
where $\lambda$ is the learning rate, $\mathcal{L}$ is the loss function, $\nabla$ is the gradient, $f_s(x)$ is the output of the surrogate model.
In our experiment, CIFAR-10 is used as the training dataset of the victim model, and ResNet20 is used as the architecture. The accuracy of the victim model is 93.30\%.
We randomly select 2,048 samples from the training dataset as the thief dataset and use the same architecture of the victim model to perform the attack. Adam is used as an optimizer with a learning rate of 0.01. The $\lambda$ is set to 0.1. The loss function is Kullback-Leibler divergence. Figure \ref{fig:Margin_JBDA} shows the training process of the surrogate model with the increasing query times.

\textbf{MAZE} \cite{kariyappa2021MAZE} generates synthetic data using a generative model without accessing any natural samples. This is achieved by a game theory optimization problem as follows:
\begin{equation}
    \min_{f_s} \max_{G} \mathbb{E}_{z \sim \mathcal{N}(0,1)} \left[ \mathcal{L}(f_v(G(z)), f_s(G(z))) \right],
\end{equation}
where \( \mathbb{E}_{z \sim \mathcal{N}(0,1)} \) denotes the expectation over the noise \( z \), which is sampled from the standard Gaussian distribution \( \mathcal{N}(0,1) \), \( \mathcal{L} \) is the loss function that measures the distance between the outputs of the victim model \( f_v \) and the surrogate model \( f_s \), \( G(z) \) is the generative model that takes noise \( z \) as input and generates samples. This optimization problem attempts to find the best surrogate model \(f_s\) while also adjusting the generative model \(G\) to produce increasingly indistinguishable samples. In our experiment, the victim model is the same as in JBDA. The adversary can access 2,048 samples, which can largely improve performance compared to that performance without any samples. The loss function is Kullback-Leibler divergence. Adam is used as the optimizer with a learning rate of 0.01 for the surrogate model and 1e-4 for the generative model. Other parameters are the same as in the original paper \cite{kariyappa2021MAZE}. Figure \ref{fig:Margin_MAME} shows the training process of the surrogate model with the increasing query times.

\section{Additional Experimental Details and Results}

\subsection{Detailed Experimental Setup}
\label{Detailed_Experimental_Setup}
We conduct experiments using eight NVIDIA RTX 4090 GPUs, each with 24GB of memory, running on an Intel(R) Xeon(R) Platinum 8352V CPU. We implemented the experiments using PyTorch 2.0.0 and Python 3.8 on an Ubuntu 20.04 system.

For victim models, we use the SGD optimizer with a learning rate of 0.01 to train the model. The momentum, weight decay, and other parameters for SGD are set to the default value. The batch size is set to 128 for CIFAR-10, CIFAR-100, FashionMNIST, and CelebA, and 32 for STL-10. The epoch of LeViT is set to 400, and the other epochs are set to 200 due to the slow convergence speed of LeViT.

For surrogate models, we use all training samples as the thief dataset for querying and training 100 epochs. The optimizer and other hyper-parameters are the same as those used in the victim models. 

In the default settings, we use 400 training samples on CIFAR-10, STL-10, FashionMNIST, and CelebA to calculate MRC. Since the category of CIFAR-100 is large, we use 40 samples on CIFAR-100 to calculate MRC. $q$ and $\eta$ are set to 0.5.

\subsection{Effectiveness of Metrics on Other Datasets}
\label{Effect_Metircs_Datasets}

\begin{figure*}[t]
    \centering
    \subfigure[FashionMNIST]{
        \includegraphics[width=0.2\textwidth]{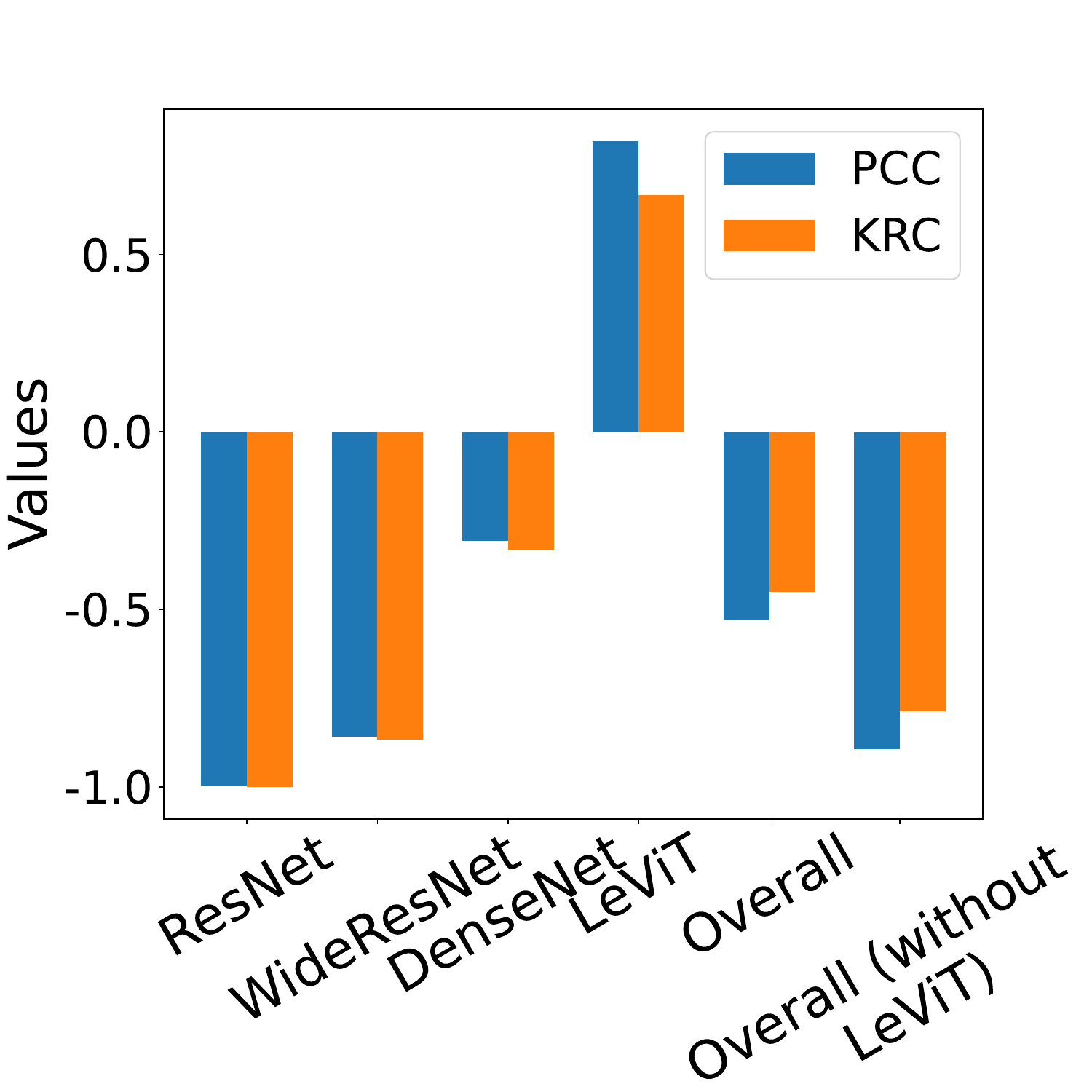}
        \label{fig:}
    }
    \subfigure[CIFAR-100]{
        \includegraphics[width=0.2\textwidth]{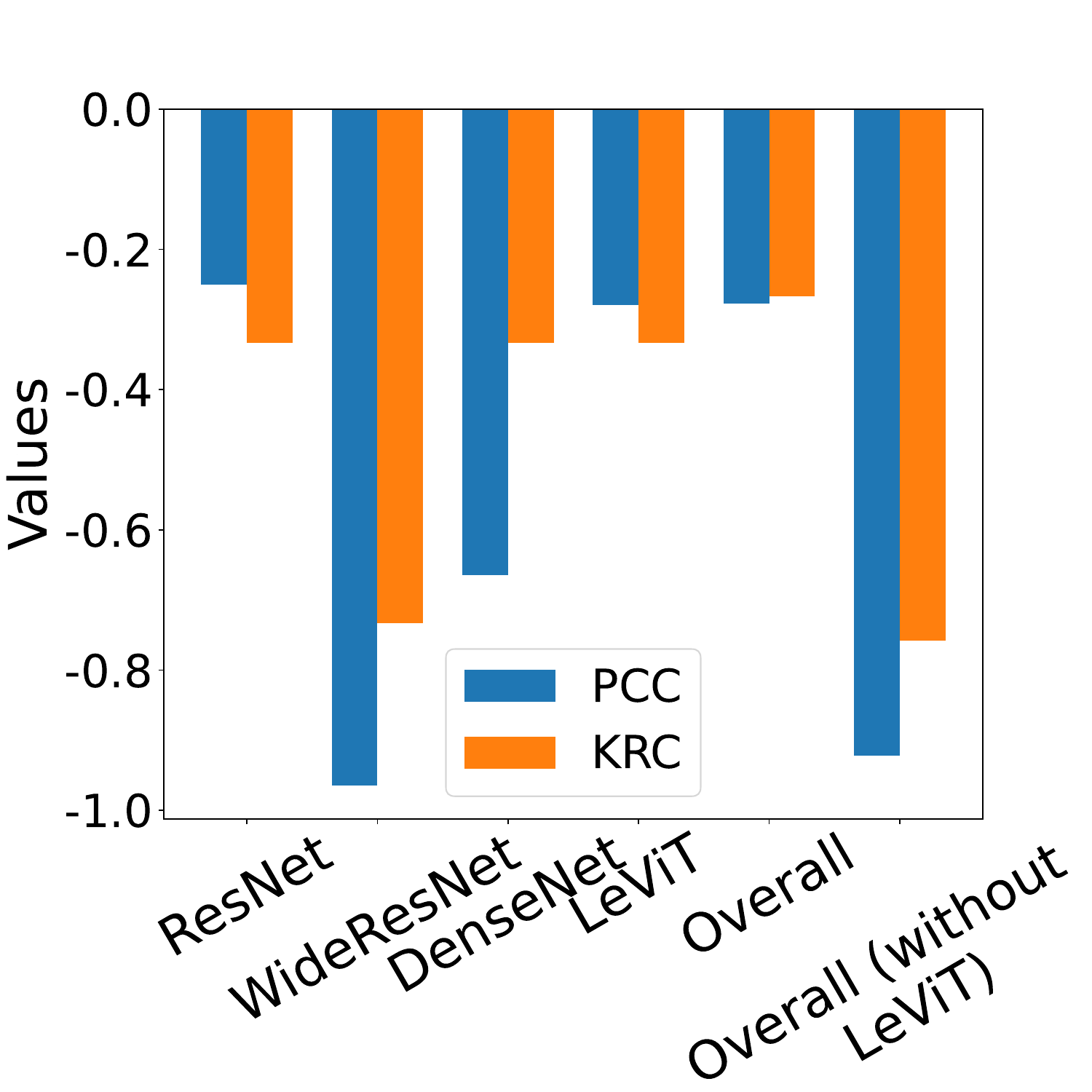}
        \label{fig:}
    }
    \subfigure[CelebA]{
        \includegraphics[width=0.2\textwidth]{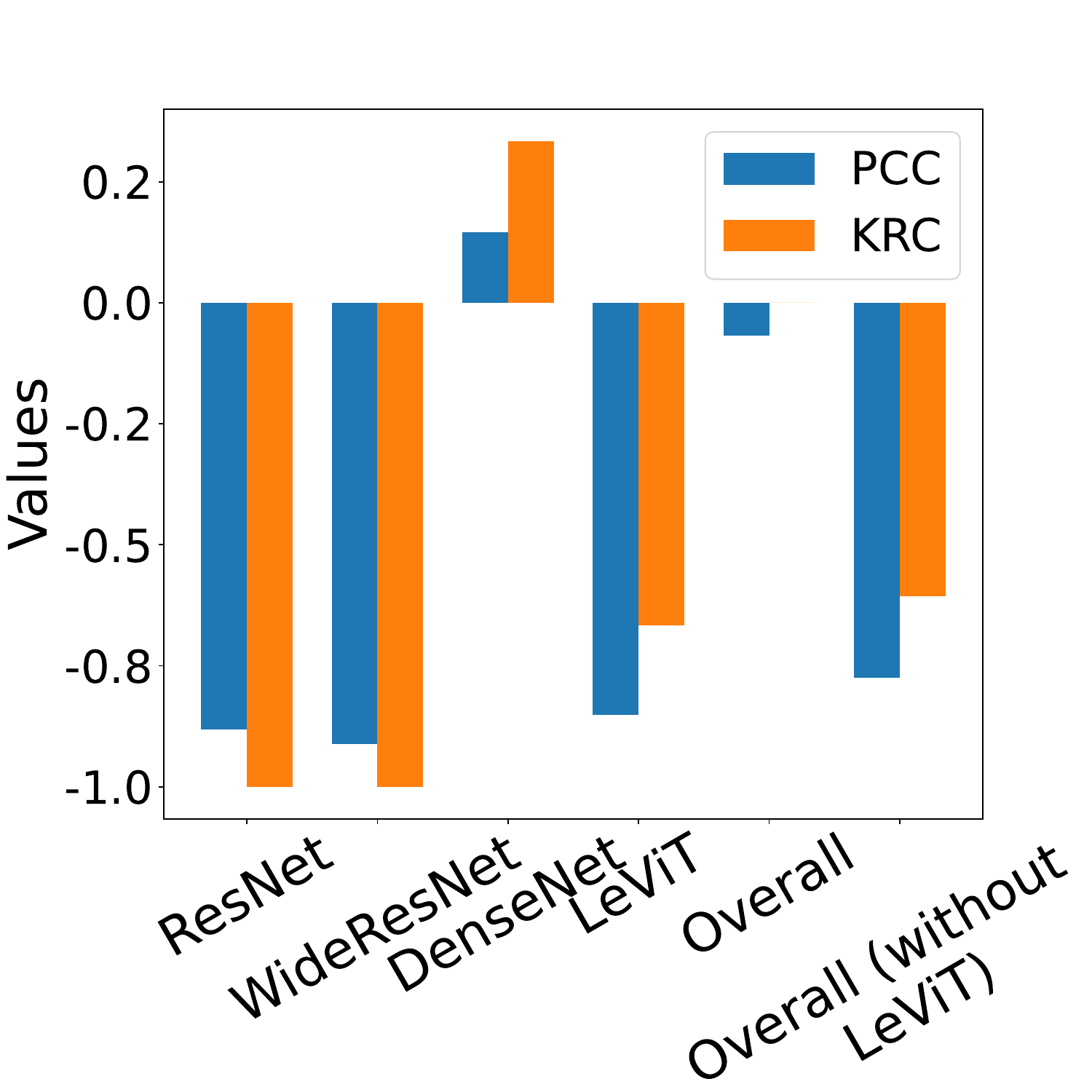}
        \label{fig:}
    }
    \caption{PCC and KRC results between MRC and attack fidelity on (a) FashionMNIST, (B) CIFAR-100 and (c) CelebA.}
    \label{fig:PCC_RC_More}
\end{figure*}

\begin{figure*}[t]
    \centering
    \subfigure[FashionMNIST]{
        \includegraphics[width=0.2\textwidth]{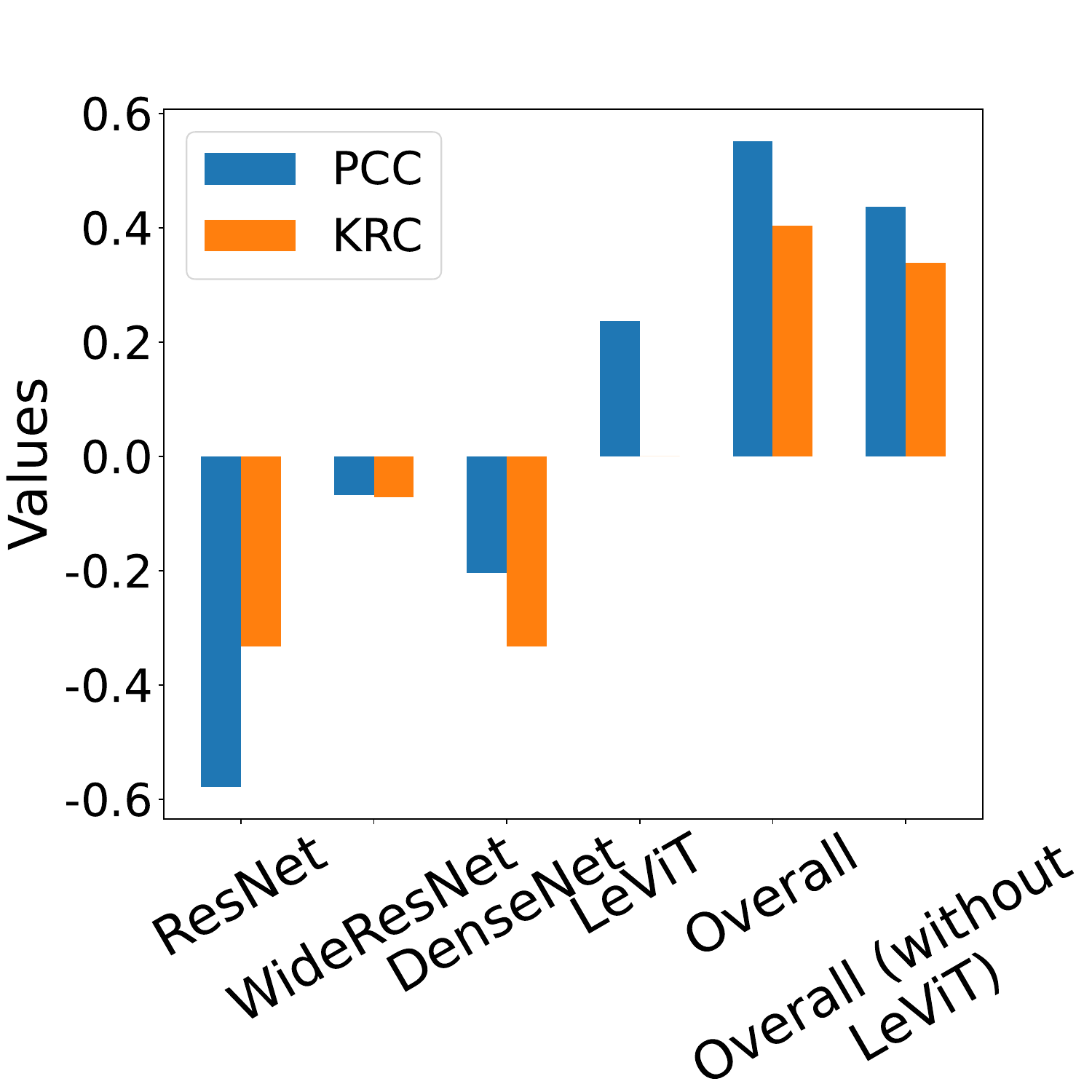}
        \label{fig:}
    }
    \subfigure[CIFAR-100]{
        \includegraphics[width=0.2\textwidth]{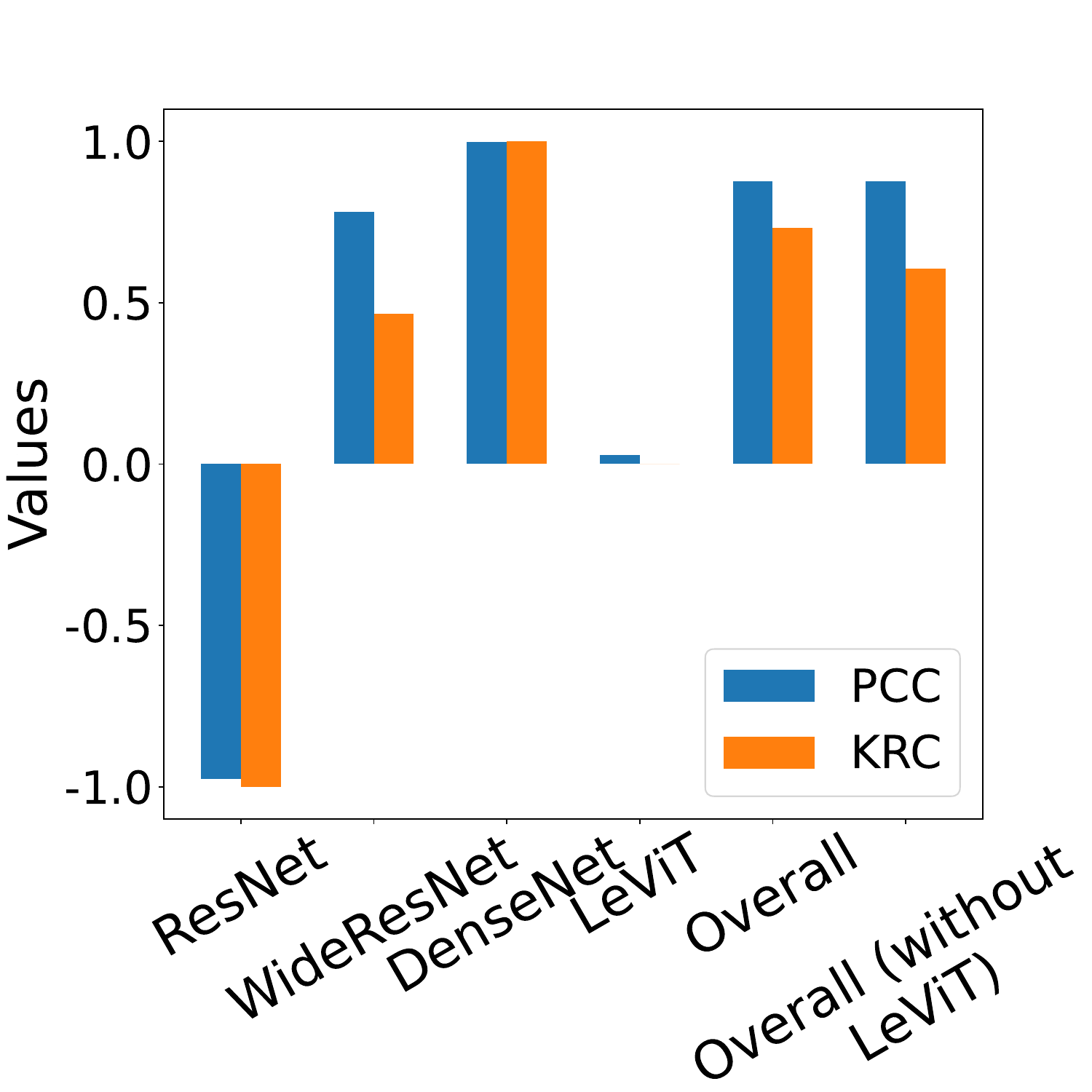}
        \label{fig:}
    }
    \subfigure[CelebA]{
        \includegraphics[width=0.2\textwidth]{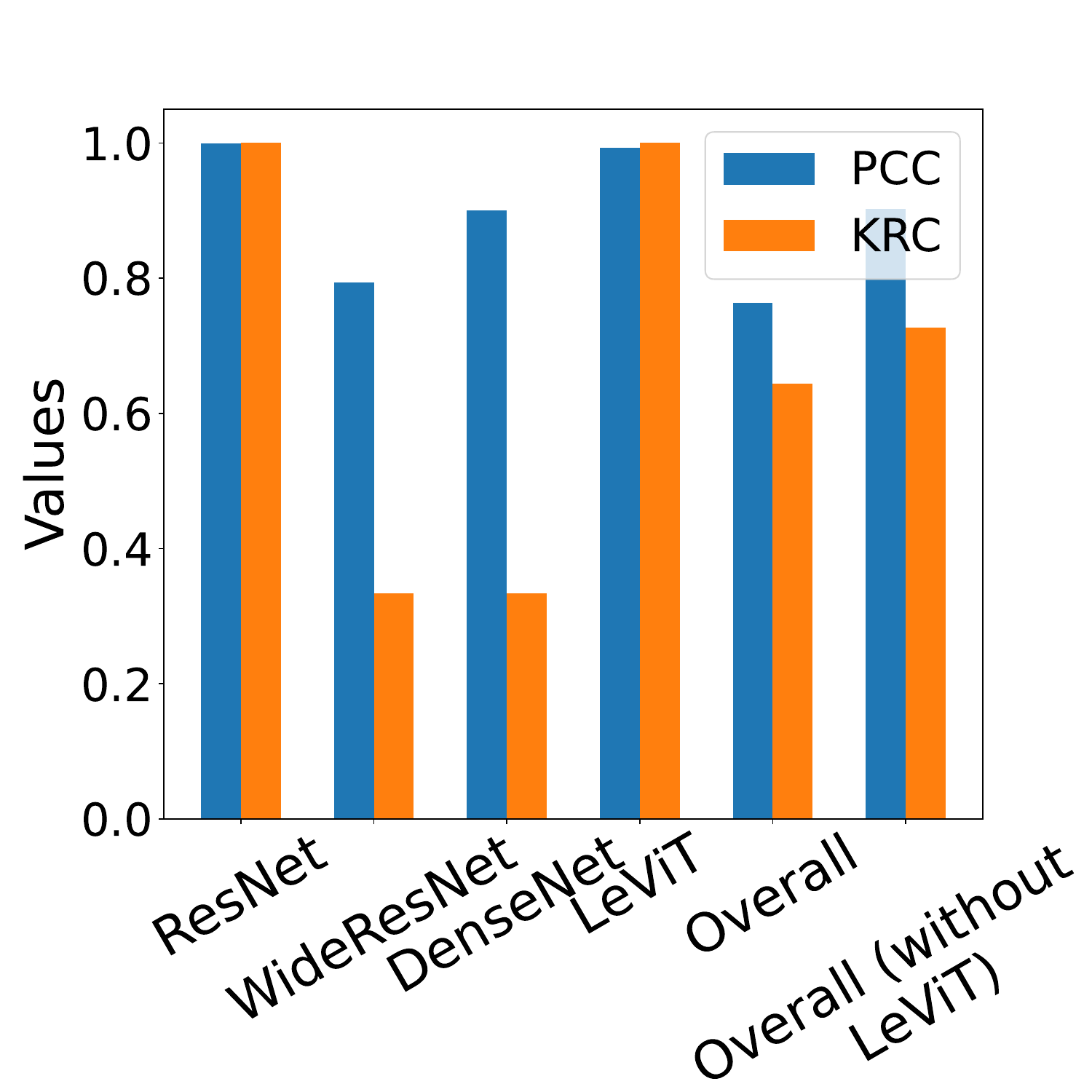}
        \label{fig:}
    }
    \caption{PCC and KRC results between VMA and attack fidelity on (a) FashionMNIST, (B) CIFAR-100 and (c) CelebA.}
    \label{fig:PCC_VMA_More}
\end{figure*}

We analyze the effectiveness of proposed metrics on more datasets.
As shown in Figure \ref{fig:PCC_RC_More}, the negative relationship between MRC and attack fidelity exists in most groups. 
There are two opposite cases: the LeViT group on FashionMNIST and the DenseNet group on CelebA. This is because the model parameters of LeViT and DenseNet are both very large. Using only 400 training samples for evaluation is insufficient, so the calculated MRC scores are inaccurate. If the number of samples used to calculate MRC increases, the accuracy of MRC scores in parameter-rich models will improve, but it will also significantly increase GPU and memory overhead. For the models with fewer parameters (ResNet, WideResNet), only 100 samples are sufficient for evaluation. Therefore, we must balance the trade-off between computational accuracy and cost consumption. In this paper, we select 400 samples to compute metrics, achieving good accuracy at an acceptable cost. In addition, Figure \ref{fig:PCC_VMA_More} further verifies that there is an overall positive relationship between VMA and attack fidelity, which is not conclusive within each group. 

With these additional results, we further validate the effectiveness of proposed metrics and also demonstrate the necessity of combining them.

\subsection{Detailed Experimental Results}
\label{Detailed_results}
\begin{table*}
	\caption{Results of CIFAR-10.}
 \vskip 0.15in
    \centering
	\label{Full_results_C10}
	\begin{tabular}{c|c|c|c|ccc}
		\toprule
        \multirow{2}{*}{\shortstack{Model}} & \multirow{2}{*}{\shortstack{\#PARA}} &\multirow{2}{*}{\shortstack{Victim Model \\Accuracy (\%)}} & \multirow{2}{*}{\shortstack{Attack Accuracy \\(\%) / Fidelity}} &\multicolumn{3}{c}{Model Recovery Complexity}\\
        
        & & & & L=100 & L=200 & L=400 \\		 
		\midrule
		ResNet20 & 272,474 & 90.94 & 80.27 / 0.8135 & 0.00106 & 0.00206 & 0.00353 \\
		ResNet32 & 466,906 & 91.66 & 80.77 / 0.8088 & 0.00101 & 0.00221 & 0.00405 \\
		ResNet44 & 661,338 & 92.41 & 79.97 / 0.8031 & 0.00106 & 0.00232 & 0.00736 \\
		WideResNet22-2 & 1,079,642 & 92.97 & 85.18 / 0.8601 & 0.00093 & 0.00168 & 0.00350 \\
		WideResNet28-2 & 1,467,610 & 93.28 & 82.86 / 0.8312 & 0.00265 & 0.00425 & 0.00797 \\
		WideResNet34-2 & 1,855,578 & 93.63 & 83.33 / 0.8365 & 0.00107 & 0.00213 & 0.00444 \\
		WideResNet40-2 & 2,243,546 & 93.46 & 82.80 / 0.8323 & 0.00250 & 0.00523 & 0.00973 \\
		WideResNet22-4 & 4,298,970 & 94.25 & 88.19 / 0.8841 & 0.00051 & 0.00095 & 0.00183 \\
		WideResNet22-8 & 17,158,106 & 94.62 & 87.98 / 0.8858 & 0.00041 & 0.00076 & 0.00135 \\
		DenseNet121 & 6,956,298 & 94.99 & 88.71 / 0.8912 & 0.00041 & 0.00061 & 0.00187 \\
		DenseNet169 & 12,493,322 & 94.67 & 88.93 / 0.8924 & 0.00041 & 0.00058 & 0.00204 \\
		DenseNet201 & 18,104,330 & 94.85 & 89.26 / 0.8969 & 0.00037 & 0.00052 & 0.00181 \\
		LeViT-128 & 8,439,386 & 81.43 & 70.97 / 0.7182 & 0.00047 & 0.00087 & 0.00180 \\
		LeViT-192 & 10,174,943 & 83.26 & 72.52 / 0.7305 & 0.00034 & 0.00065 & 0.00130 \\
		LeViT-256 & 17,865,166 & 83.62 & 73.69 / 0.7347 & 0.00027 & 0.00053 & 0.00110 \\
            LeViT-384 & 37,586,862 & 84.00 & 72.30 / 0.7255 & 0.00026 & 0.00057 & 0.00131 \\
\bottomrule
\end{tabular}
\vskip 0.15in
\end{table*}

\begin{table*}
	\caption{Results of STL-10.}
 \vskip 0.15in
        \centering
	\label{Full_results_STL10}
	\begin{tabular}{c|c|c|c|ccc}
		\toprule
             \multirow{2}{*}{\shortstack{Model}} & \multirow{2}{*}{\shortstack{\#PARA}} &\multirow{2}{*}{\shortstack{Victim Model \\Accuracy (\%)}} & \multirow{2}{*}{\shortstack{Attack Accuracy \\(\%) / Fidelity}} &\multicolumn{3}{c}{Model Recovery Complexity}
            \\

		 &  &  &  &  L=100& L=200 & L=400\\		 
		\midrule
        ResNet20 & 4,327,754 & 83.96 & 74.10 / 0.7699 & 0.00337 & 0.00802 & 0.01811 \\
        ResNet32 & 7,427,914 & 83.79 & 72.39 / 0.7430 & 0.00334 & 0.00857 & 0.01939 \\
        ResNet44 & 10,528,074 & 84.54 & 70.90 / 0.7300 & 0.00390 & 0.00901 & 0.02142 \\
        WideResNet22-2 & 1,079,642 & 83.89 & 66.33 / 0.6770 & 0.00594 & 0.01507 & 0.03317 \\
        WideResNet28-2 & 1,467,610 & 85.28 & 65.96 / 0.6705 & 0.00729 & 0.01668 & 0.03602 \\
        WideResNet34-2 & 1,855,578 & 85.19 & 66.08 / 0.6761 & 0.00563 & 0.01311 & 0.02709 \\
        WideResNet40-2 & 2,243,546 & 85.25 & 64.53 / 0.6679 & 0.00882 & 0.01708 & 0.03666 \\
        WideResNet22-4 & 4,298,970 & 85.61 & 67.45 / 0.7011 & 0.00365 & 0.00903 & 0.02032 \\
        WideResNet22-8 & 17,158,106 & 87.28 & 72.01 / 0.7474 & 0.00330 & 0.00725 & 0.01654 \\
        DenseNet121 & 6,956,298 & 85.33 & 82.00 / 0.8451 & 0.00278 & 0.00637 & 0.02611 \\
        DenseNet169 & 12,493,322 & 87.40 & 83.29 / 0.8536 & 0.00244 & 0.00481 & 0.02415 \\
        DenseNet201 & 18,104,330 & 86.47 & 82.82 / 0.8479 & 0.00233 & 0.00484 & 0.02563 \\
        LeViT-128 & 8,440,682 & 68.51 & 48.20 / 0.4810 & 0.00327 & 0.00787 & 0.02251 \\
        LeViT-192 & 10,175,823 & 72.55 & 45.20 / 0.4626 & 0.00302 & 0.00892 & 0.02835 \\
        LeViT-256 & 17,866,318 & 73.14 & 45.45 / 0.4612 & 0.00327 & 0.01001 & 0.03054 \\
        LeViT-384 & 37,588,574 & 72.10 & 46.30 / 0.4818 & 0.00392 & 0.01134 & 0.02965 \\
		\bottomrule
	\end{tabular}
\vskip 0.15in
\end{table*}

\begin{table*}
	\caption{Results of FashionMNIST.}
 \vskip 0.15in
        \centering
	\label{Full_results_F10}
	\begin{tabular}{c|c|c|c|ccc}
		\toprule
             \multirow{2}{*}{\shortstack{Model}} & \multirow{2}{*}{\shortstack{\#PARA}} &\multirow{2}{*}{\shortstack{Victim Model \\Accuracy (\%)}} & \multirow{2}{*}{\shortstack{Attack Accuracy \\(\%) / Fidelity}} &\multicolumn{3}{c}{Model Recovery Complexity}
            \\

		 &  &  &  &  L=100& L=200 & L=400\\		 
		\midrule
		ResNet20 & 272,826 & 93.68 & 92.16 / 0.9383 & 0.00103 & 0.00235 & 0.00398 \\
        ResNet32 & 467,258 & 94.13 & 91.98 / 0.9365 & 0.00198 & 0.00339 & 0.00632 \\
        ResNet44 & 661,690 & 94.07 & 91.39 / 0.9303 & 0.00280 & 0.00698 & 0.01226 \\
        WideResNet22-2 & 1,079,354 & 94.67 & 90.50 / 0.9381 & 0.00057 & 0.00105 & 0.00262 \\
        WideResNet28-2 & 1,467,322 & 94.84 & 92.02 / 0.9301 & 0.00228 & 0.00391 & 0.00677 \\
        WideResNet34-2 & 1,855,290 & 94.85 & 92.15 / 0.9355 & 0.00132 & 0.00257 & 0.00482 \\
        WideResNet40-2	& 2,243,258 & 95.01 & 91.35 / 0.9310 & 0.00397 & 0.00671 & 0.01145 \\
        WideResNet22-4 & 4,298,682 & 95.01& 91.42 / 0.9427 & 0.00057 & 0.00108 & 0.00216 \\
        WideResNet22-8 & 17,157,818 & 94.84 & 91.34 / 0.9449 & 0.00050 & 0.00084 & 0.00142 \\
        DenseNet121 & 6,955,146 & 95.27 & 93.52 / 0.9432 & 0.00051 & 0.00082 & 0.00084 \\
        DenseNet169 & 12,492,170 & 95.21 & 92.96 / 0.9482 & 0.00052 & 0.00082 & 0.00078 \\
        DenseNet201 & 18,103,178 & 95.17 & 92.04 / 0.9437 & 0.00049 & 0.00076 & 0.00076 \\
        LeViT-128 & 8,439,098 & 92.09 & 91.06 / 0.9339 & 0.00047 & 0.00098 & 0.00233 \\
        LeViT-192 & 10,174,511 & 93.03 & 91.52 / 0.9379 & 0.00045 & 0.00093 & 0.00204 \\
        LeViT-256 & 17,864,590 & 92.61 & 90.96 / 0.9324 & 0.00035 & 0.00072 & 0.00168 \\
        LeViT-384 & 37,585,998 & 92.62 & 90.78 / 0.9252 & 0.00028 & 0.00059 & 0.00126 \\

\bottomrule
\end{tabular}
\vskip 0.15in
\end{table*}

\begin{table*}
	\caption{Results of CIFAR-100.}
 \vskip 0.15in
        \centering
	\label{Full_results_C100}
	\begin{tabular}{c|c|c|c|cc}
		\toprule
             \multirow{2}{*}{\shortstack{Model}} & \multirow{2}{*}{\shortstack{\#PARA}} &\multirow{2}{*}{\shortstack{Victim Model \\Accuracy (\%)}} & \multirow{2}{*}{\shortstack{Attack Accuracy \\(\%) / Fidelity}} &\multicolumn{2}{c}{Model Recovery Complexity}
            \\

		 &  &  &  &  L=20& L=40 \\		 
		\midrule
        ResNet20 & 278,324 & 65.76 & 51.37 / 0.5498 & 0.01945 & 0.09119 \\
        ResNet32 & 472,756 & 66.79  & 50.54 / 0.5156 & 0.02615 & 0.13944 \\
        ResNet44 & 667,188 & 68.25  & 48.17 / 0.4931 & 0.03096 & 0.09816 \\
        WideResNet22-2 & 1,091,252 & 69.22  & 53.85 / 0.5457 & 0.02527 & 0.06880 \\
        WideResNet28-2 & 1,479,220 & 69.40  & 57.06 / 0.5707 & 0.03112 & 0.07286 \\
        WideResNet34-2 & 1,867,188 & 70.53  & 49.88 / 0.5042 & 0.03050 & 0.08139 \\
        WideResNet40-2 & 2,255,156 & 71.03  & 50.40 / 0.5050 & 0.03940 & 0.09775 \\
        WideResNet22-4 & 4,322,100 & 73.26  & 60.81 / 0.6221 & 0.01274 & 0.03498 \\
        WideResNet22-8 & 17,204,276 & 76.14  & 64.08 / 0.6579 & 0.00909 & 0.01826 \\
        DenseNet121 & 7,048,548 & 76.13  & 65.03 / 0.6633 & 0.00453 & 0.00009 \\
        DenseNet169 & 12,643,172 & 77.15  & 70.71 / 0.7266 & 0.00353 & 0.00005 \\
        DenseNet201 & 18,277,220 & 77.22  & 71.40 / 0.7353 & 0.00390 &
        0.00008 \\
        LeViT-128 & 8,474,036 & 48.35 & 38.33 / 0.4164 & 0.00312 & 0.00565 \\
        LeViT-192 & 10,209,593 & 50.69 & 40.07 / 0.4321 & 0.00558 & 0.01247 \\
        LeViT-256 & 17,911,336 & 54.57 & 39.84 / 0.4064 & 0.00261 & 0.02479 \\
        LeViT-384 & 37,656,072 & 56.68 & 41.32 / 0.4294 & 0.00875 & 0.02019 \\
\bottomrule
\end{tabular}
\vskip 0.15in
\end{table*}

\begin{table*}
	\caption{Results of CelebA.}
 \vskip 0.15in
        \centering
	\label{Full_results_C8}
	\begin{tabular}{c|c|c|c|ccc}
		\toprule
             \multirow{2}{*}{\shortstack{Model}} & \multirow{2}{*}{\shortstack{\#PARA}} &\multirow{2}{*}{\shortstack{Victim Model \\Accuracy (\%)}} & \multirow{2}{*}{\shortstack{Attack Accuracy \\(\%) / Fidelity}} &\multicolumn{3}{c}{Model Recovery Complexity}
            \\

		 &  &  &  &  L=100& L=200 & L=400\\		 
		\midrule
ResNet20 & 272,344 & 71.62  & 69.81 / 0.7437 & 0.02541 & 0.04658 & 0.09466 \\
        ResNet32 & 466,776 & 71.17  & 71.29 / 0.7610 & 0.01486 & 0.02508 & 0.04914 \\
        ResNet44 & 661,208 & 72.02  & 69.61 / 0.7533 & 0.02453 & 0.05230 & 0.08990 \\
        WideResNet22-2 & 1,079,384 & 73.77  & 70.57 / 0.7650 & 0.04911 & 0.06883 & 0.12100 \\
        WideResNet28-2 & 1,467,352 & 73.55  & 70.87 / 0.7801 & 0.02911 & 0.04271 & 0.06296 \\
        WideResNet34-2 & 1,855,320 & 72.82  & 71.99 / 0.7783 & 0.03305 & 0.04729 & 0.06558 \\
        WideResNet40-2 & 2,243,288 & 72.82  & 71.94 / 0.7703 & 0.03751 & 0.05336 & 0.07329 \\
        WideResNet22-4 & 4,298,456 & 74.05  & 73.45 / 0.8104 & 0.01138 & 0.01429 & 0.02040 \\
        WideResNet22-8 & 17,157,080 & 74.67  & 73.72 / 0.8324 & 0.00562 & 0.00803 & 0.01143 \\
        DenseNet121 & 6,954,248 & 73.90  & 72.65 / 0.7938 & 0.03182 & 0.05440 & 0.07745 \\
        DenseNet169 & 12,489,992 & 74.37  & 71.22 / 0.7963 & 0.01396 & 0.01844 & 0.02806 \\
        DenseNet201 & 18,100,488 & 74.00  & 72.85 / 0.7913 & 0.01107 & 0.01437 & 0.01891 \\
        LeViT-128 & 8,439,048 & 67.23  & 70.79 / 0.7282 & 0.00211 & 0.00382 & 0.00773 \\
        LeViT-192 & 10,174,476 & 69.94  & 71.64 / 0.7485 & 0.00103 & 0.00249 & 0.00509 \\
        LeViT-256 & 17,864,536 & 68.74  & 70.22 / 0.7402 & 0.00129 & 0.00224 & 0.00413 \\
        LeViT-384 & 37,585,912 & 69.44  & 69.61 / 0.7523 & 0.00119 & 0.00193 & 0.00363 \\
        \bottomrule
    \end{tabular}
    \vskip 0.15in
\end{table*}
We give specific numerical values of the results, including the number of parameters (denoted as \#PARA), VMA, attack fidelity, attack accuracy, and the calculated MRC. The particular results for CIFAR-10 can be found in Table \ref{Full_results_C10}. Similarly, the results for STL-10 in Table \ref{Full_results_STL10}, for FashionMNIST in Table \ref{Full_results_F10}, for CIFAR-100 are detailed in Table \ref{Full_results_C100}, and for CelebA, the data is provided in Table \ref{Full_results_C8}. 
Intuitively, \#PARA may be related to model complexity, thus reflecting the difficulty of MEAs. We experimentally verify the correlation between \#PARA and attack fidelity. Contrary to expectations, the results indicate that no definite relationship exists between \#PARA and attack fidelity, with a PCC of -0.28 and a KRC of 0.15 on CIFAR-10.


\subsection{Detailed Results on Impact of $\eta$}
\label{detailed_eat}
\begin{table*}
	\caption{Results on the impact of the difficulty ratio $\eta$.}
 \vskip 0.15in
        \centering
	\label{Detailed_eta}
	\begin{tabular}{c|cccccc}
		\toprule
             \multirow{2}{*}{\shortstack{Model}}  &\multicolumn{6}{c}{Model Recovery Complexity}
            \\
		 & $\eta=0$ & $\eta=0.25$ & $\eta=0.5$ & $\eta=0.75$ & $\eta=1$	& Random\\ 
		\midrule
        ResNet20 & \textbf{0.00278} & 0.00324 & 0.00353 & 0.00388 & 0.00416 & 0.00729\\
        ResNet32 & 0.00532 & 0.00528 & \textbf{0.00405} & 0.00465 & 0.00569 & 0.00734\\
        ResNet44 & \textbf{0.00231} & 0.00408 & 0.00736 & 0.00698 & 0.00839 & 0.00620\\
        WideResNet22-2 &\textbf{ 0.00127} & 0.00230 & 0.00350 & 0.00455 & 0.00504 & 0.00295\\
        WideResNet28-2 & \textbf{0.00464} & 0.00645 & 0.00797 & 0.00949 & 0.01002 & 0.00949\\
        WideResNet34-2 & \textbf{0.00286} & 0.00357 & 0.00444 & 0.00533 & 0.00601 & 0.00654\\
        WideResNet40-2 & \textbf{0.00364} & 0.00665 & 0.00973 & 0.01157 & 0.01182 & 0.00971\\
        WideResNet22-4 & \textbf{0.00054} & 0.00127 & 0.00183 & 0.00228 & 0.00247 & 0.00134\\
        WideResNet22-8 & \textbf{0.00049} & 0.00102 & 0.00135 & 0.00151 & 0.00136 & 0.00103\\
        DenseNet121 & 0.00193 & 0.00205 & 0.00187 & 0.00180 & \textbf{0.00124} & 0.00212\\
        DenseNet169 & 0.00177 & 0.00201 & 0.00204 & 0.00184 & \textbf{0.00136} & 0.00214\\
        DenseNet201 & 0.00160 & 0.00172 & 0.00181 & 0.00161 & \textbf{0.00110} & 0.00201\\
        LeViT-128 & \textbf{0.00090} & 0.00138 & 0.00180 & 0.00213 & 0.00238 & 0.00527\\
        LeViT-192 & \textbf{0.00085} & 0.00109 & 0.00130 & 0.00149 & 0.00166 & 0.00336\\
        LeViT-256 & \textbf{0.00067} & 0.00088 & 0.00110 & 0.00135 & 0.00174 & 0.00235\\
        LeViT-384 & \textbf{0.00055} & 0.00084 & 0.00131 & 0.00183 & 0.00241 & 0.00172\\
\bottomrule
\end{tabular}
\vskip 0.15in
\end{table*}
		
We analyze the impact of difficulty ratio $\eta$ on CIFAR-10, and the detailed results are given in Table \ref{Detailed_eta}. It is worth noting that the MRC scores obtained from randomly selected samples are occasionally lower than those computed solely from hard samples ($\eta = 1$). This observation elucidates why attack strategies relying on randomly selected samples can occasionally outperform active learning-based approaches~\cite{pal2020ActiveThief,tifrea2023margin}.
We also observe that with 400 samples, many models have lower MRC scores when $\eta = 0$. This emphasizes the importance of simple samples, especially when the sample size is small. This conclusion aligns with the pruning strategy proposed by \citet{sorscher2022Neural}, suggesting the use of simple samples in data-scarce scenarios.
In our results, it is observed that only within the DenseNet group with the best generalization ability, the lowest MRC score is obtained when $\eta=1$, which means that hard samples hold greater significance when attempting to extract models with superior generalization capabilities.

\end{document}